\documentclass{jfm}
\usepackage{bm,amsmath,amssymb,natbib,graphicx,mathrsfs,float}
\usepackage{array,epic,eepic,color,upgreek,psfrag,cleveref,cancel}

\graphicspath{{./Figs/}}

\newcommand{\W}{Wi}

\newcommand{\R}{Re}

\newlength{\piclen}
\title[Vortex merging and splitting events in viscoelastic Taylor Couette flow. ]
{Vortex merging and splitting events in viscoelastic Taylor Couette flow}

\author[Jose M. Lopez] 
{J\ls O\ls
  S\ls E\ns M.\ns L\ls O\ls P\ls E\ls Z}

\affiliation{Universitat Polit\`ecnica de Catalunya, Physics Department \\
Campus Nord UPC, 08034 Barcelona, Spain\\[\affilskip] }

\pubyear{} \volume{} \pagerange{}
\date{\today}
\begin{document}

\maketitle

\begin{abstract}

Recent experiments have reported a novel transition to elasto-inertial turbulence in the 
Taylor--Couette flow of a dilute polymer solution. Unlike previously reported transitions, 
this newly discovered scenario, dubbed \emph{vortex merging and splitting} (VMS) transition, 
occurs in the centrifugally unstable regime and the mechanisms underlying it 
are two-dimensional: the flow becomes chaotic due to the proliferation of 
events where axisymmetric vortex pairs may be either created (vortex splitting) or 
annihilated (vortex merging). 
In this paper, we present direct numerical simulations, 
using the FENE-P constitutive equation to model polymer dynamics, which reproduce 
the experimental observations with great accuracy and elucidate the reasons for 
the onset of this surprising dynamics. Starting from the Newtonian limit and 
increasing progressively the fluid's elasticity, we demonstrate that 
the VMS dynamics is not associated with the well-known Taylor vortices, 
but with a steady pattern of elastically induced axisymmetric vortex pairs 
known as diwhirls. The amount of angular momentum carried by these elastic 
vortices becomes increasingly small as the fluid's elasticity increases 
and it eventually reaches a marginal level. When this occurs, the diwhirls 
become dynamically disconnected from the rest of the system and move independently 
from each other in the axial direction. It is shown that vortex merging and splitting 
events, along with local transient chaotic dynamics, result from the interactions 
among diwhirls, and that this complex spatio-temporal dynamics persists even 
at elasticity levels twice as large as those investigated experimentally.

\end{abstract}

\section{Introduction}

\noindent The Taylor--Couette flow (TCF) i.e. the fluid flow contained in the annular gap between two vertical concentric cylinders, 
is a prototypical system to investigate hydrodynamic instabilities and turbulence in rotating flows. If the working fluid 
is Newtonian and only the inner cylinder is rotated, this system provides one of the best known examples of a supercritical transition 
to the turbulence~\citep{coles65,FensSwiGol79}. When the rotation speed exceeds a critical value, the initial purely azimuthal flow, 
known as circular Couette flow (CCF), becomes unstable due to a centrifugal instability, leading to a stationary axisymmetric pattern 
of toroidal vortices known as Taylor vortex flow (TVF) \citep{Taylor23}. 
As the rotation speed increases, the TVF is gradually replaced by flows of increasing spatio-temporal complexity, giving rise eventually to a fully turbulent state. 
The characteristics of the transitions and flow regimes that precede the onset of turbulence  depend on the geometry of the system (i.e. the curvature and the length-to-gap aspect ratio). 
However, when the curvature is moderate, as is the case in most experiments, the route to chaos involves a series of Hopf bifurcations, i.e. the so-called Ruelle-Takens scenario~\citep{RueTa71}. 
The physical mechanisms underlying these transitions, as well as the spatial and temporal properties of the pre-turbulent flow regimes, 
have been widely investigated over decades and are now  relatively well understood~\citep{jones_1981,gorman_swinney_1982,king_lee_swinney_marcus_1984,marcus_1984_1,marcus_1984_2,jones_1985,andereck_liu_swinney_1986,HegBaxAnd96,wereley_lueptow_1998,MaSerrLuep14,DeTuWeBaWi18}. 

\noindent This archetypal transition scenario, as well as the properties of the eventual turbulent state, are however substantially modified when long chain polymers (even in small amounts) are added to the working fluid. Unlike Newtonian fluids, the response of dilute polymer solutions to the flow stresses is not only a function of the strain, but also of the strain rate. This time-dependent behaviour of the fluid properties, known as viscoelasticity, often causes dramatic changes in the stability and spatio-temporal characteristics of the flow with respect to those of the Newtonian case. The most striking manifestation of this phenomenon is the occurrence of flow instability in the absence of inertia~\citep{muller1989purely,larson_shaqfeh_muller_1990}. This instability results from the combined effect of elasticity and curvature and produces different flow regimes depending on the elasticity level of the working fluid. When the elasticity level is moderate, the instability leads to a steady vortex pattern similar to the TVF~\citep{baumert1995flow,BauMu97,AlSuKho99}. However, when the solution is highly elastic, the flow exhibits a form of chaotic motion dubbed elastic turbulence~\citep{GroStei00,GroStei04}. 

\noindent In parameter regimes where inertial effects are not negligible, the interplay between elasticity and inertia leads to a rich variety of flow patterns and spatio-temporal behaviours. The regions of existence of the different flow regimes in the parameter space defined by the elasticity level and the rotation speed (normally quantified by the dimensionless Elasticity and Reynolds numbers, $El$ and $Re$, respectively) are very sensitive to the experimental protocols and the polymers properties. Particularly significant among these properties is the shear thinning behaviour of the dilute polymer solution. Recent experiments have shown that strong shear thinning may even fully suppress elasto-inertially induced flow regimes~\citep{cagney_lacassagne_balabani_2020,lacassagne_cagney_balabani_2021}. In cases where elastic effects prevail over shear thinning effects (e.g. Boger-like fluids), experiments and simulations have reported a number of flow regimes. These can be roughly divided into coherent and chaotic flow states. The most characteristic examples of coherent states are the so-called Ribbons (RB), diwhirls (DW) and oscillatory strips (OS)~\citep{GroStei96,GroiStei97,BAUMERT99,CruMuGrise02,ThoSurKho06,thomas_khomami_sureshkumar_2009}. The RB arise from a supercritical Hopf bifurcation of the CCF at low $Re$ values and consist in a rotating standing wave pattern formed by  two spiral waves propagating axially in opposite senses. In contrast, DW and OS emerge from non-linear instabilities (in many cases as secondary instabilities of the RB pattern) and are vortex pairs characterized by strong inflows, which are confined within narrow axial regions, and weak outflows, which extend over axial distances that are usually three or four times larger than those of the inflows. The difference between DW and OS is that whereas the former is stationary, the latter is oscillatory. Both these structures have the ability to merge when they are close to each other and usually appear as spatially localized states~\citep{GroiStei97}. The regimes of chaotic motion can be achieved either from the non-linear development of these coherent flow patterns or directly from CCF via subcritical transition. The most typical examples of chaotic dynamics are disorder oscillations, spatio-temporal intermittency and flames-like turbulence~\citep{GroStei96,ThoSurKho06,BauMu97,BAUMERT99,ThoSurKho06,thomas_khomami_sureshkumar_2009,LaCruMu12,liu_khomami_2013}. Ultimately, when the rotation speed becomes sufficiently large, the flow reaches a state of highly disordered motion involving a  wide range of spatial and temporal scales. This state is known as elasto-inertial turbulence (EIT) and it exhibits structural and statistical features which are markedly distinct from those of ordinary Newtonian turbulence~\citep{liu_khomami_2013,song_teng_liu_ding_lu_khomami_2019,song_lin_liu_lu_khomami_2021}.

\noindent  In recent years, there has been a surge of interest in investigating the distinct pathways followed by the flow to achieve the 
EIT state. Experimental studies have so far identified three main types of transition scenarios. In the first type, dubbed \emph{defect-mediated} (DM) transition~\citep{LaCruMu12,LaAbCruMu16}, the route to EIT starts when the state of RB undergoes a Benjamin-Fair instability which produces spatio-temporal defects in the flow pattern. The number of defects grows as $Re$ increases, creating increasingly large regions of irregular spatio-temporal behaviour. This results first in a regime of spatio-temporal intermittency and subsequently in a fully chaotic flow that was identified as EIT. The DM transition takes place at low-to-moderate $El$ values ($El < 0.15$) and $Re$ values quite below those at which the onset of TVF happens in the Newtonian case. The second type of transition, known as transition via \emph{flames} (F)~\citep{latrache2021transition}, occurs at similar $Re$ but larger $El$ values ($0.15 \leq El \leq 0.3$). Again, the transition is initiated from the RB pattern, which in this case undergoes an instability that results in the emergence of flame-like structures. These flames proliferate as the rotation speed increases, increasing progressively the spatio-temporal complexity of the flow until the EIT regime is achieved. The third transition scenario is dubbed \emph{vortex merging and splitting} (VMS) transition~\citep{LaCaGiBa20}. Unlike the two previous transitions, the VMS occurs at $Re$ values where CCF is centrifugally unstable and the primary instability results in a steady axisymmetric vortex flow that the authors identified as a TVF. Here, the spatio-temporal complexity of the flow increases following a temporal sequence of events in which the vortex pairs may be either annihilated or created. The frequency with which these events occur increases with increasing $Re$, giving rise eventually to a highly chaotic dynamics consistent with a EIT state. 

\noindent While the experiments have shown that this vortex merging and splitting dynamics is of elastic nature~\citep{LaCaGiBa20}, the reasons why axisymmetric vortex pairs undergo merging and splitting processes and why they occur at relatively high $El$ levels are not known. This paper aims to shed some light on these aspects. Numerical simulations of viscoelastic TCF, using the FENEP model to simulate polymers effects, are used to study the progressive transformation that the vortex flow pattern undergoes as $El$ increases from the Newtonian limit ($El = 0$) up to $El$ values well beyond the onset of the VMS dynamics. In contrast to what was thought, the simulations reveal that the VMS dynamics is not associated with a centrifugally driven TVF-like pattern, but with an elastically induced pattern of steady DW that fully replaces the TVF pattern at $El \approx 0.12$, where the instability mechanism changes from being centrifugal to being elastic. These elastic vortices have a striking feature that had not been previously reported: the amount of angular momentum they carry decreases with increasing $El$.  It is shown that the VMS dynamics starts when $El$ is sufficiently large so that the contribution of these vortices to the angular momentum flux reaches a marginal level. The distinct vortex pairs become then dynamically disconnected from the rest of the system and begin to travel independently in the axial direction, creating the complex spatio temporal dynamics characteristic of the VMS regime.

\section{Problem formulation, dimensionless parameters and numerical methods}\label{sec:Problem}

\noindent We consider the motion of a dilute polymer solution confined to the gap between two vertical, rigid and concentric cylinders of height $h$ and radii $r_i$ and $r_o$.  Hereafter, the subscripts $i$ and $o$ denote the inner and outer cylinders, respectively. The inner cylinder rotates with an angular velocity, $\Omega_i$, whereas the outer cylinder is at rest, i.e. $\Omega_o = 0$. The dynamics of this incompressible viscoelastic fluid flow is governed by the continuity and Navier-Stokes equations, along with an equation to describe the temporal evolution of a polymer conformation tensor, $\mathbf{C}$, which contains the ensemble average elongation and orientation of all polymer molecules in the flow. A simple Hookean dumbbell model is used to represent the polymer molecules~\citep{Bird80}. Normalizing the velocity with  the inner cylinder velocity, $\Omega_i r_i$, the length with the gap size,  $d=r_o - r_i$, the pressure with the dynamic pressure, $\rho (\Omega_i r_i)^2$, where $\rho$ is the fluid's density, and the polymer conformation tensor with $kT_e/H$, where $k$ denotes the Boltzmann constant, $T_e$ is the absolute temperature and $H$ is the spring constant, the dimensionless equations read
\begin{equation}\label{eq:gov_eq}
    \begin{aligned}
      \nabla \cdot \mathbf{v} = 0,\\
      \partial_t\mathbf{\mathbf{v}} + \mathbf{v}\cdot\nabla\mathbf{v} = 
      -\nabla P + \frac{\beta}{Re} \nabla^2\mathbf{v} + \frac{(1-\beta)}{Re} \nabla\cdot \mathbf{T},\\
      \partial_t\mathbf{C}+\mathbf{v}\cdot\nabla\mathbf{C} =
      \mathbf{C}\cdot\nabla\mathbf{v} + (\nabla\mathbf{v})^T\cdot\mathbf{C}- \mathbf{T},
    \end{aligned}
\end{equation}
where $\mathbf{v}=(u,v,w)$ is the velocity vector field in cylindrical coordinates $(r,\theta,z)$, $P$ is the pressure, $\beta = \nu_s/\nu$ indicates the relative importance between the solvent viscosity $\nu_s$ and
the viscosity of the solution at zero shear rate $\nu$ and $Re= \Omega_i r_i d/\nu$ is the Reynolds number based on the inner cylinder velocity. Polymers are coupled to the Navier-Stokes equations through the polymer stress tensor $\mathbf{T}$, which is calculated using the FENEP model~\citep{Bird80},
\begin{equation}\label{Peterlin}
  \mathbf{T} = \frac{1}{\W}(\frac{\mathbf{C}}{1-\frac{tr(\mathbf{C})}{L^2}}-\mathbf{I}),
\end{equation}
where $\mathbf{I}$ is the unit tensor, $tr(\mathbf{C})$ is the trace of the polymer conformation tensor, $L$ denotes the maximum polymer extension and $Wi$ is the Weissenberg number, a dimensionless quantity that measures the ratio of the polymer relaxation time $\lambda$ to the advective time scale  $d/(\Omega_i r_i)$. \\

\noindent Experimental observations in~\cite{LaCaGiBa20} strongly suggest that the dynamics relevant to the vortex 
merging and splitting transition are two-dimensional and occur in the meridional plane ($r,z$). Based on this 
assumption, the simulations were conducted in a quasi-2D TC system (i.e. under axisymmetric conditions), 
where the velocity field does not depend on the azimuthal coordinate, $\theta$. This choice allows us to significantly reduce the cost of the simulations, making it possible to simulate the viscoelastic flow for very long periods of time. Some simulations in a fully 3D TC system were also subsequently performed to verify that the  dynamics found in the axisymmetric simulations persist in full domain.\\

\noindent Periodic boundary conditions are used in the $z$ direction, whereas the dimensionless boundary conditions at the cylinders are 
\begin{equation}\label{eq:bc_cylinders}
    \begin{aligned}
      \mathbf{v}(r_i,z) = (0,1,0),\\
      \mathbf{v}(r_o,z) = (0,0,0).
      \end{aligned}
\end{equation}
The parameters used in the simulations have been chosen to mimic as closely as possible those in the experiments of~\cite{LaCaGiBa20}. The curvature of the system is the same as in the experiments, $\eta = r_i/r_o = 0.77$, and all simulations have been performed at a constant value of the Reynolds number, $Re = 95$, consistent with the $Re$ value at which the onset of complex spatio-temporal dynamics takes place in the experiments. It must be noted that at this $Re$ value, the laminar Couette flow is centrifugally unstable, and therefore the flow pattern in the Newtonian case consists of Taylor vortices (the onset of Taylor vortices occurs at $Re=89$ for this value of $\eta$). The same concentration of polymers as in the experiments has been used, $\beta = 0.871$, and similar levels of the polymers elasticity, quantified by the elasticity number, $El = Wi/Re$, have also been considered.
 
 There are, however, other parameters and features that could not be matched.  The height-to-gap aspect ratio, $\Gamma = h/d$, in the simulations had to be reduced with respect to that in the experiments, $\Gamma = 21.5$, to keep the computational cost affordable. The majority of the simulations have been performed using $\Gamma = 12.6$. Simulations at other values of $\Gamma$, spanning between $9$ and $16$, have also been conducted to assess the influence of 
 this parameter on the dynamics observed in the simulations (section~\ref{sec:domain_size_and_others}). Another difference with respect to the experimental setup is the absence of endplates. While secondary flows resulting from the interaction between flow and endplates are known to often alter the stability properties and dynamics of Newtonian TCF~\citep{CzSeBoLue03,AvGriLoMa08,LoAv17},~\cite{LaCaGiBa20} noted that this does not seem to be the case in their experiments. Hence, simulations in an axially periodic domain are expected to provide a good qualitative representation of the observed dynamics. Finally, other parameter that sets a difference with respect to the experiments is the maximum polymer extension $L$. This parameter of the FENEP model is a property of the dilute polymer solution which cannot be easily inferred from the specifications of the polymer used in the experiments (polyacrylamide, Sigma-Aldrich,$M_w = 5.5 \times 10^6$ g/mol).  Most of the simulations presented in this paper have been conducted using $L = 100$, which is a standard value in the literature of viscoelastic TC flows~\citep{liu_khomami_2013,song_teng_liu_ding_lu_khomami_2019,song_lin_liu_lu_khomami_2021}. The influence 
 of varying this parameter is analysed in the section~\ref{sec:domain_size_and_others}, where simulations with $L$ varying between $30$ and $250$ are presented and discussed.\\

\noindent To solve the equations~\eqref{eq:gov_eq}, we use our open source code nsCouette~\citep{lopez2020nscouette}, 
which has been recently extended to simulate viscoelastic flows using the FENEP model. 
This code is a highly scalable pseudo-spectral solver for annular flows that implements a very 
efficient hybrid parallelization strategy~\citep[see][for details]{shi2015hybrid}. 
The spatial discretization in the $z$ direction is carried out via Fourier-Galerkin expansion, 
whereas high order central finite differences on a Gauss-Lobatto-Chebyshev grid are used in $r$. 
A Pressure Poisson Equation (PPE) formulation is used to decouple the pressure from the velocity field. 
The free divergence condition (i.e. the continuity 
equation) is enforced by using an influence matrix technique, so that this condition is satisfied up to machine error. 
The time integration was carried out using a second order accurate predictor-corrector scheme based 
on the Crank-Nicolson method~\citep{openpipeflow}. Further details about the timestepper can be found in~\cite{lopez_choueiri_hof_2019}. 
As customary in numerical simulations of viscoelastic flows using pseudospectral codes, a small amount of artificial diffusion 
is added to stabilize the integration. The necessity to include this diffusion arises from the hyperbolic 
nature of the time evolution equation for $\mathbf{C}$. The absence of a diffusive term in this equation leads to 
an accumulation of numerical error that in many cases results in numerical breakdown. To avoid this problem, 
a Laplacian term, $\frac{1}{\R S_c} \nabla^2 \mathbf{C}$, is added to the right hand side of that equation, 
where the Schmidt number, $S_c=\nu/\kappa$, quantifies the ratio between the viscous and artificial diffusivities. 
In the simulations presented here, $S_c=100$, which yields an artificial diffusion coefficient, $\frac{1}{Re S_c} = 10^{-4}$. 
This amount of diffusion is similar in magnitude to those used in other recent numerical studies on 
viscoelastic flows~\citep{XiGra10b,lopez_choueiri_hof_2019,song_teng_liu_ding_lu_khomami_2019,zhu_song_liu_lu_khomami_2020,song_wan_liu_lu_khomami_2021,zhu_song_lin_liu_lu_khomami_2022}. It has been verified that a reduction in the amount of diffusion  does not significantly alter the results of the simulations ($Sc$ numbers up to $200$ were tested), thereby confirming the adequacy of the diffusion used for the simulations throughout the paper. Due to the addition of a Laplacian term, boundary conditions for $\mathbf{C}$ are needed at the cylinders. To avoid introduction of artificial boundary conditions, the values of $\mathbf{C}$ at the cylinders are obtained by solving its evolution equation without the artificial diffusion term. This strategy was first introduced by~\cite{Beris99} and it has been widely used since then. The number of radial nodes and Fourier modes used in the computations are shown in table~\ref{res} for the different values of $\Gamma$ considered. The time step size had to be varied between $4\cdot10^{-3}$ and $10^{-3}$ as the polymers elasticity (i.e. the $El$ number) was increased.

\begin{table}
  \begin{center}
    \begin{tabular}{c c c}
      $\Gamma$ & $m_r$ & $m_z$  \\\hline
       9 & 128  & 192 \\
      10 & 128  & 256 \\
      12.6 & 128  & 256 \\
      14 & 128  & 384 \\
      16 & 128  & 512 \\\hline
    \end{tabular}
  \end{center}
  \caption{Number of radial nodes ($m_r$) and axial Fourier modes ($m_z$) used in the simulations depending on the aspect ratio $\Gamma$ of the system.}
 \label{res}
\end{table}

\section{Results}\label{sec:to_VMST}

\subsection{Transition to the VMS regime with increasing $El$}\label{sec:to_VMST}

\noindent We first investigate the gradual approach to the VMS regime as the elasticity of the fluid increases. For this initial simulation, 
an aspect ratio of $\Gamma=12.56$ was considered, whereas the maximum polymer extension was set to $L = 100$. A Newtonian simulation (i.e. $\beta = 1$)  was initially run to calculate a Taylor vortex flow pattern with six pairs of counter rotating vortices. Using this state as initial condition, the fluid's elasticity was slowly and steadily increased at a uniform rate, $El = 10^{-3}t/Re$, until a dynamical regime characterized by merging and splitting events was found. We would like to stress that the protocol followed in this  simulation differs from that in the experiments, where the VMS regime is achieved by increasing $Re$ while keeping a constant elasticity~\citep{LaCaGiBa20}. Our study hence offers a different perspective into the pathway leading to this flow regime and allows to identify the gradual transformation the flow undergoes as the working fluid becomes more elastic.\\

\begin{figure}
      \includegraphics[width=\linewidth]{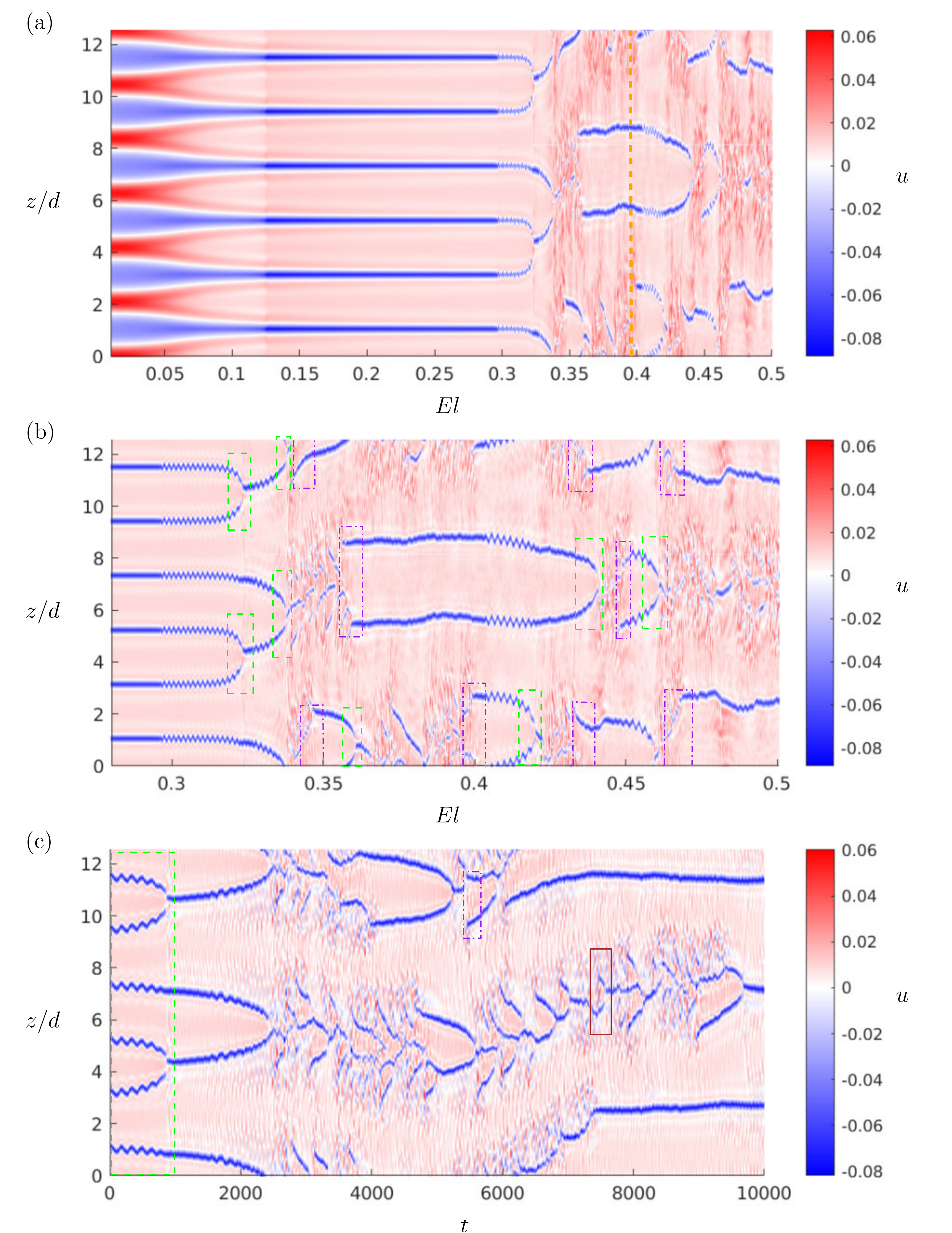} 
  \caption{ (Color online) Space-time plot representing the magnitude of the radial velocity $u$ at mid-gap along the axial direction $z$.  Panels (a) and (b) correspond to a simulation performed at $Re = 95$ where, starting from a Newtonian Taylor vortex flow, $El$ was slowly increased with time ($El = 10^{-3}t/Re$). Note that the time $t$ has been replaced by the corresponding $El$ values in the horizontal axis. Panel (a) shows the variation of $u$ from Newtonian flow (i.e. $El = 0$) up to the largest $El$ value simulated ($El = 0.50$), whereas panel (b) shows in more detail the range of $El$ values for which complex spatio-temporal dynamics take place. Panel (c) illustrates the spatio-temporal dynamics when the $El$ number is kept constant after the VMS regime is achieved. The case exemplified corresponds to $El = 0.32$. Red and blue areas indicate outflows and inflows respectively. Note that periodic boundary conditions are used in $z$.}
  \label{fig:spacetime_El}
\end{figure}

\noindent Panel (a) in figure~\ref{fig:spacetime_El} provides an overview of the structural and dynamical changes induced by the polymers on the initial Taylor vortex flow as $El$ increases. It shows a space-time diagram of the radial velocity $u$ at mid-gap along the axial direction $z$, where time has been replaced by its corresponding $El$ values. Red areas represent fluid motion from the inner to the outer cylinder, i.e. outflows, whereas blue regions indicate fluid moving from the outer to the inner cylinder, i.e. inflows. Panel (b) places emphasis in the range of $El$ values for which complex spatio-temporal dynamics happens. The stationary pattern of vortices becomes unstable at $El \approx 0.29$ leading to periodic oscillations of the vortex pairs along the $z$-axis. 
The onset of the VMS regime takes place soon after, at $El \approx 0.315$, as the dynamics of the distinct vortex pairs decouple and these begin to move independently in the axial direction. It will be shown later in section~\ref{sec:domain_size_and_others} that this threshold is sensitive to the number of vortices of the initial condition and the aspect ratio used in the simulations. Merging events, where two vortex pairs coalesce to form a single vortex pair, are indicated as dashed (green) rectangles in the panel (b) of the figure~\ref{fig:spacetime_El}. These events fully dominate the dynamics in the initial phase of the VMS regime (for $0.32 < El < 0.34$) and since they occur simultaneously at different axial locations, the total number of vortex pairs in the system rapidly decreases. After this initial phase ($El > 0.34$), merging events coexist with events where a vortex pair branches into two, i.e. splitting events, shown as dash-dotted (purple) rectangles in the figure, as well as with regions where the dynamics becomes transiently chaotic (see for instance the flow  region between $4 < z/d < 7$ for $0.34 < El < 0.36$ or $0 < z/d < 2.5$ for $0.38 < El < 0.40$). The number of vortex pairs fluctuates between two and four in this phase.\\ 

\noindent  It is important to note that the occurrence of VMS events does not depend on the continuous increase of $El$ with time. If $El$ is held constant after the VMS regime is achieved, the simulations show the same dynamic events just described: vortex merging, vortex splitting and transient chaotic motion, reflecting that these are temporal characteristics of the flow that occur when $El$ exceeds a certain critical threshold. This is demonstrated in the panel (c) of the figure~\ref{fig:spacetime_El}, which shows a space time map for a simulation where $El$ has been fixed to $0.32$. 
Interestingly, the VMS events observed in simulations with constant $El$ are similar to those observed in simulations where $El$ varies with time. The reason (which will be discussed later in the paper) is that increasing $El$ has little influence on the vortices in this flow regime. As a result, space time diagrams corresponding to simulations where $El$ changes with time not only feature the various flow regimes obtained when $El$ is varied but also provide an accurate representation of the VMS events.\\

\noindent Another important feature that is clearly illustrated in the panel (a) of the figure~\ref{fig:spacetime_El} is the strong impact that increasing $El$ has on the structure of the vortex pairs. We anticipate here that these structural changes are key to understanding the physics underlying merging and splitting events. Hence, before getting into detail about the dynamics in the VMS regime, it is convenient to present a comprehensive study about the influence of elasticity on the Taylor vortex flow.

\subsection{Viscoelastic modification of the Taylor vortex flow}\label{sec:VTVF}

A well known property of viscoelastic flows with curved streamlines is the appearance of a radially inward force which is caused by the elastic stresses arising from the stretching of the polymer molecules by the primary flow~\citep{GroiStei98}. This force has been identified as the driving source of a number of instabilities in curvilinear flows of highly elastic polymer solutions, which are usually known as purely elastic instabilities~\citep{Sha96}. The mechanism underlying these instabilities has been discussed in detail and verified in many studies, particularly in flow regimes where inertial effects are negligible ($Re \to 0$). It is however reasonable to expect that this elastic force will also have an influence in parameter regimes where Newtonian flows become unstable due to inertial forces. The stationary pattern of Newtonian Taylor vortices used as starting solution in our calculations is one such case: it arises from a centrifugal instability of the purely azimuthal primary flow~\citep{Taylor23}. This instability mechanism is expected to persist in the viscoelastic case as $El$ is slowly increased starting from the Newtonian limit. However, the structure of the Taylor vortex pattern is likely to be modified by the competition between the centrifugal and elastically induced forces. Additionally, if the fluid's elasticity becomes sufficiently large, the elastic instability mechanism might replace the centrifugal mechanism, leading to a flow state that is elastic in nature but whose structure could be modified by the presence of inertial effects. In this section we show that this is indeed the case in our simulations.\\

\begin{figure}
      \includegraphics[width=\linewidth]{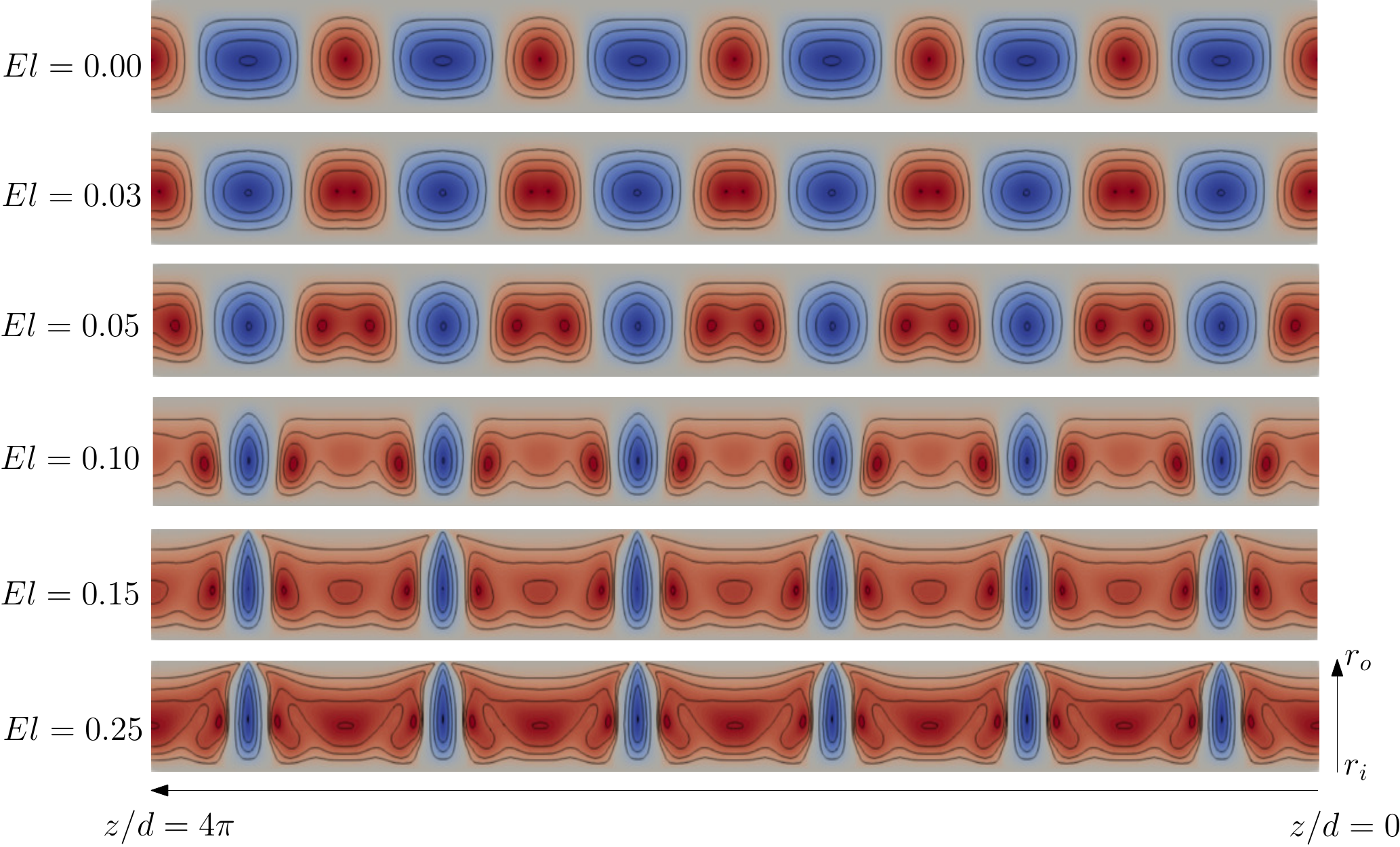} 
  \caption{ (Color online) Structural variation of the Taylor vortex flow pattern as $El$ increases from the Newtonian case (i.e. $El = 0.00$) up to values near the onset of spatio-temporal dynamics. Each panel shows a colormap of the radial velocity $u$ in a meridional plane $(z/d,r/d) \in [0,4\pi] \times [3.35,4.35]$, where red regions indicate outflows (i.e. positive velocity), blue regions represent inflows (i.e. negative velocity) and the zero velocity has been set as gray. The color scale is based on the maximum and minimum values of each case and hence differs among the different panels. There are 4 positive and 4 negative contours evenly distributed across the full range of values in each case: 1. $El = 0.00$, $u$ in $[-0.037,0.059]$; 2. $El = 0.03$, $u$ in $[-0.043,0.047]$; 3. $El = 0.05$, $u$ in $[-0.048,0.029]$; 4. $El = 0.10$, $u$ in $[-0.048,0.014]$; 5. $El = 0.15$, $u$ in $[-0.082,0.015]$; 6. $El = 0.25$, $u$ in $[-0.072,0.011]$. The system is shown rotated by 90 degrees in the counterclockwise direction with respect to its original position. The location of the inner and outer cylinders, $r_i$ and $r_o$, respectively, as well as the locations of the top ($z/d = 4\pi$) and bottom ($z/d =0$) of the system are indicated in the bottom panel.} 
  \label{fig:cmap_radial_velocity}
\end{figure}

\noindent Figure~\ref{fig:cmap_radial_velocity} shows colormaps of the radial velocity, $u$, illustrating the dependence of the flow structure as $El$ increases from the Newtonian limit ($El = 0$) up to the regime in which the flow exhibits spatio-temporal behaviour. Note that to save space, in all figures illustrating flow patterns throughout the paper, the system is shown rotated by $90$ degrees in the counterclockwise direction, so that the inner (outer) cylinder is located at the bottom (top) of each panel and the 
positive $z$-direction goes from right to left (see the  coordinate system in the bottom panel of figure~\ref{fig:cmap_radial_velocity}). The structure of the Taylor vortex flow pattern in the Newtonian case (top panel) shows a small asymmetry between outflows and inflows, as the axial extent of the inflows is slightly greater than that of the outflows. This characteristic fully reverses as elasticity comes into play. The axial extent of the inflows decreases with increasing $El$ and these become eventually confined to strong jets that extend over narrow regions in the axial direction. Conversely, the axial extent of the outflows increases notably (note that they become nearly four times larger than the inflows for $El \geq 0.12$) and the magnitude of $u$ in these regions decreases substantially as $El$ increases (the range of values of $u$ corresponding to each panel is specified in the caption).\\

\begin{figure}
      \includegraphics[width=\linewidth]{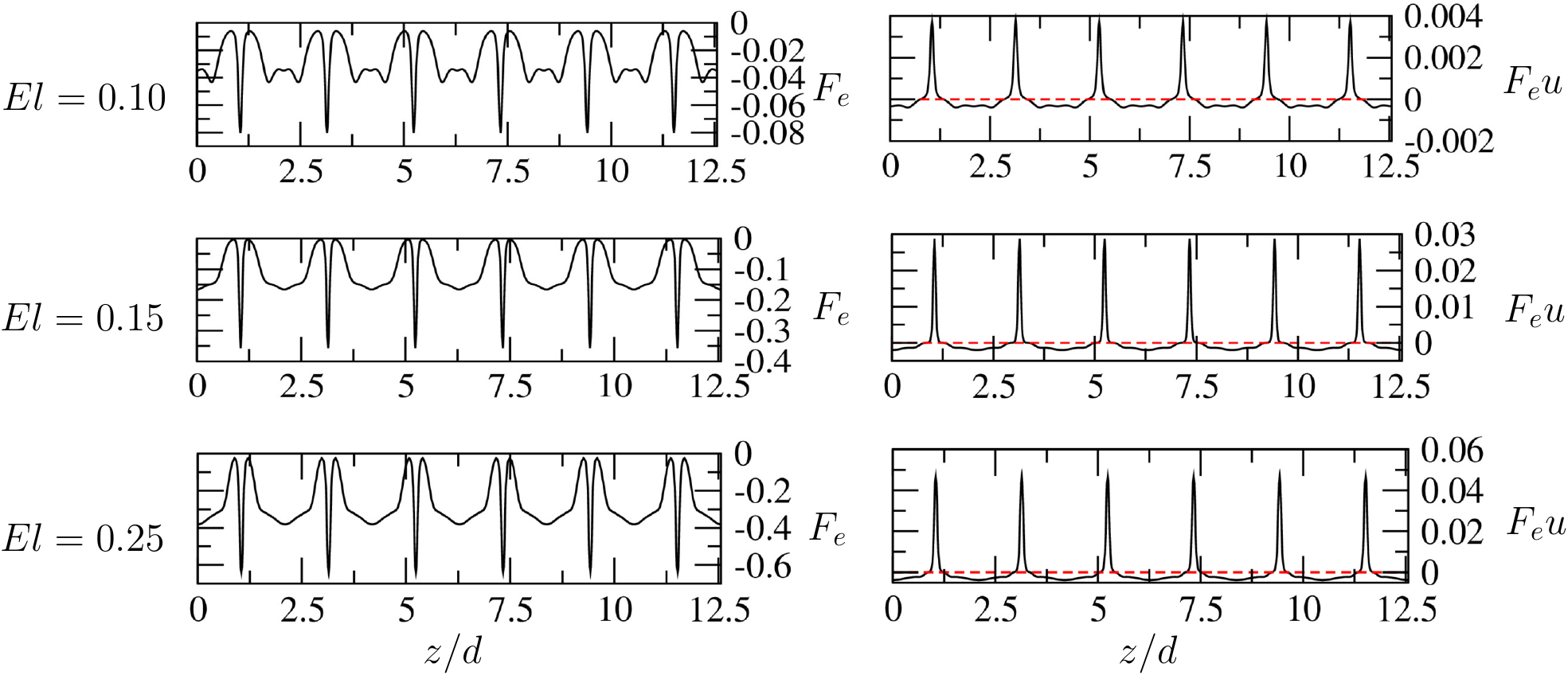} 
  \caption{ Axial profiles of the elastic force $F_e$ (left panels) and its associated work $F_e u$ (right panels) obtained at the mid-gap for $El$ values matching those of the last three panels in figure~\ref{fig:cmap_radial_velocity}. The elastic force is calculated as $F_e = \frac{(1-\beta)}{Re} (\partial_r T_{rr} + \frac{(T_{rr}-T_{\theta\theta})}{r} + \partial_z T_{rz})$. A dashed line has been added at $F_e u = 0$ to help identify inflows ($F_e u > 0$) and outflows ($F_e u < 0$).}
  \label{fig:pol_work}
\end{figure}

\noindent  It was postulated in a previous experimental study that this strong asymmetry between inflows and outflows might be caused by the work done by the elastically induced force~\citep{GroiStei98}. To verify this hypothesis quantitatively, figure~\ref{fig:pol_work} shows axial profiles of the elastic force, hereafter denoted as $F_e$ (left panels), and its associated work $F_e u$ (right panels), obtained at the mid-gap for the last three cases shown in the figure~\ref{fig:cmap_radial_velocity}. As seen, the profiles of $F_e$ are always negative, reflecting that $F_e$ is an inward force, and they exhibit strong peaks in the inflows whose magnitude increases with increasing $El$. Since $F_e$ acts in the same direction as $u$ in the inflows, it does positive work on the flow in these regions. This circumstance implies that the strong peaks of $F_e$ will result in large positive work (see the peaks of $F_e u$ in the right panels), which enhances the fluid motion in the inflows and create the strong localized jets that appear as $El$ increases. In the outflows, on the contrary, $F_e$ acts in opposition to the fluid motion and therefore does a negative work on the flow. This characteristic explains the decay in the magnitude of $u$ that is observed in the outflows as $El$ increases. The axial extent of the inward jets decreases as the magnitude of the peaks grows, which evidently entails an increase in the axial extent of the outflows and creates the asymmetry between inflows and outflows observed in figure~\ref{fig:cmap_radial_velocity}.\\

\noindent In addition to the emergence of this asymmetry, a second transformation takes place inside the outflows. The region where the largest positive velocity occurs, which in the Newtonian case is located at the centre of the outflows, separates in the viscoelastic cases into two identical regions which are symmetric with respect to the central symmetry plane of the outflow. These new regions of maximum positive velocity move away from each other as $El$ increases and approach progressively the adjacent inflows. When the elasticity is sufficiently large ($El \geq 0.12$), the strong inflow jets are flanked by these regions of maximum positive velocity, whereas the flow in the central part of the outflows becomes nearly uniform in the axial direction.\\ 

\begin{figure}\setlength{\piclen}{\linewidth}
  \begin{center}
    \includegraphics[width=\piclen]{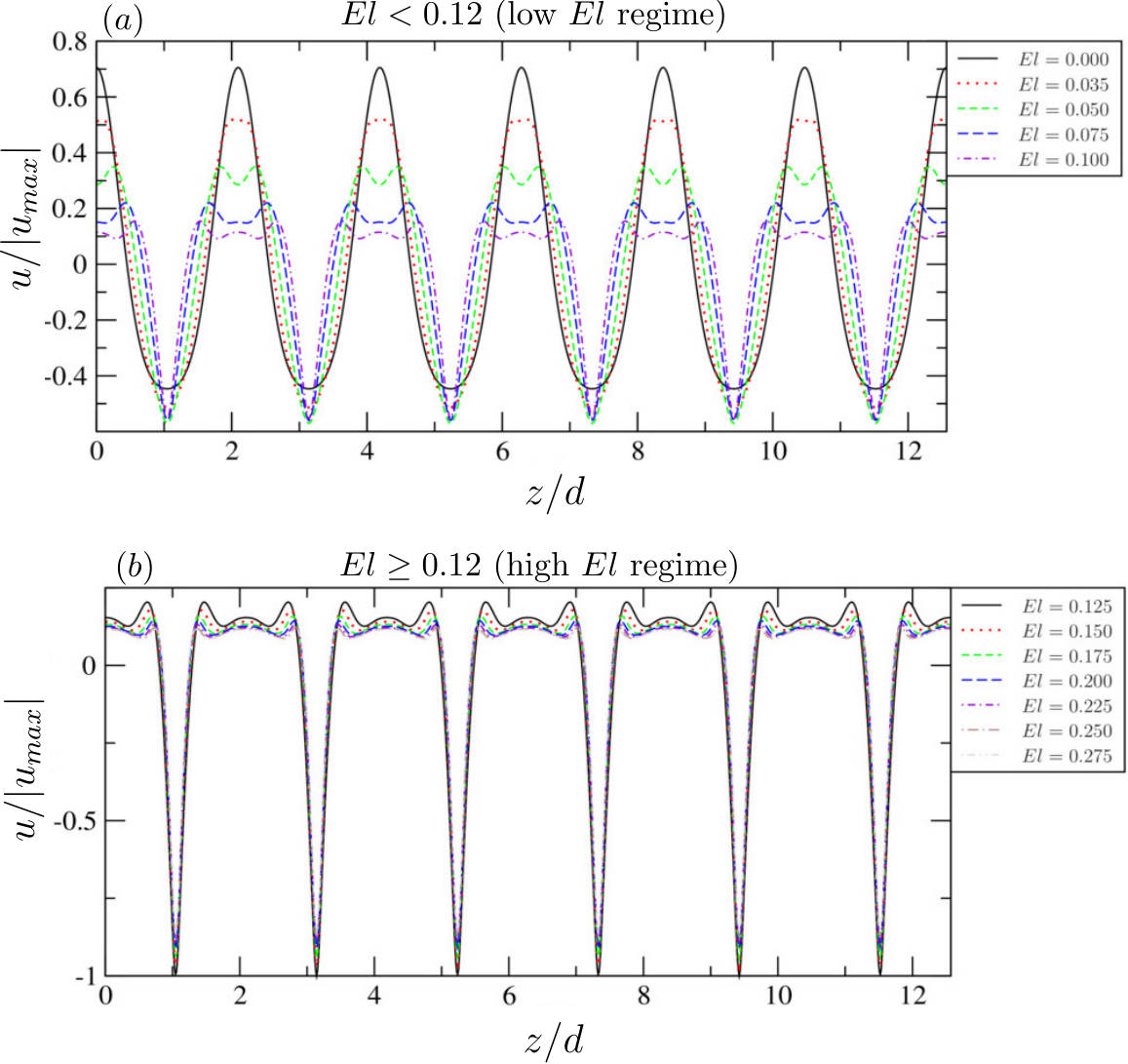}
  \end{center}
  \caption{ (Color online) Profiles of the radial velocity $u$ at mid-gap along the $z$-axis. Panel $(a)$ shows profiles for $El < 0.12$, where profound changes in the structure of the flow pattern are observed in figure~\ref{fig:cmap_radial_velocity}, whereas panel $(b)$ focuses on the range of $El$ values 
  ($El \geq 0.12$) where the variation in the profiles is small. Note that in both panels $u$ is normalized with the largest absolute value among all cases 
  which is obtained for $El = 0.12$ and corresponds to $|u_{max}| = 0.0824$.}
  \label{fig:prof_rad_vel}
\end{figure}

\begin{figure}\setlength{\piclen}{0.95\linewidth}
  \begin{center}
    \includegraphics[width=\piclen]{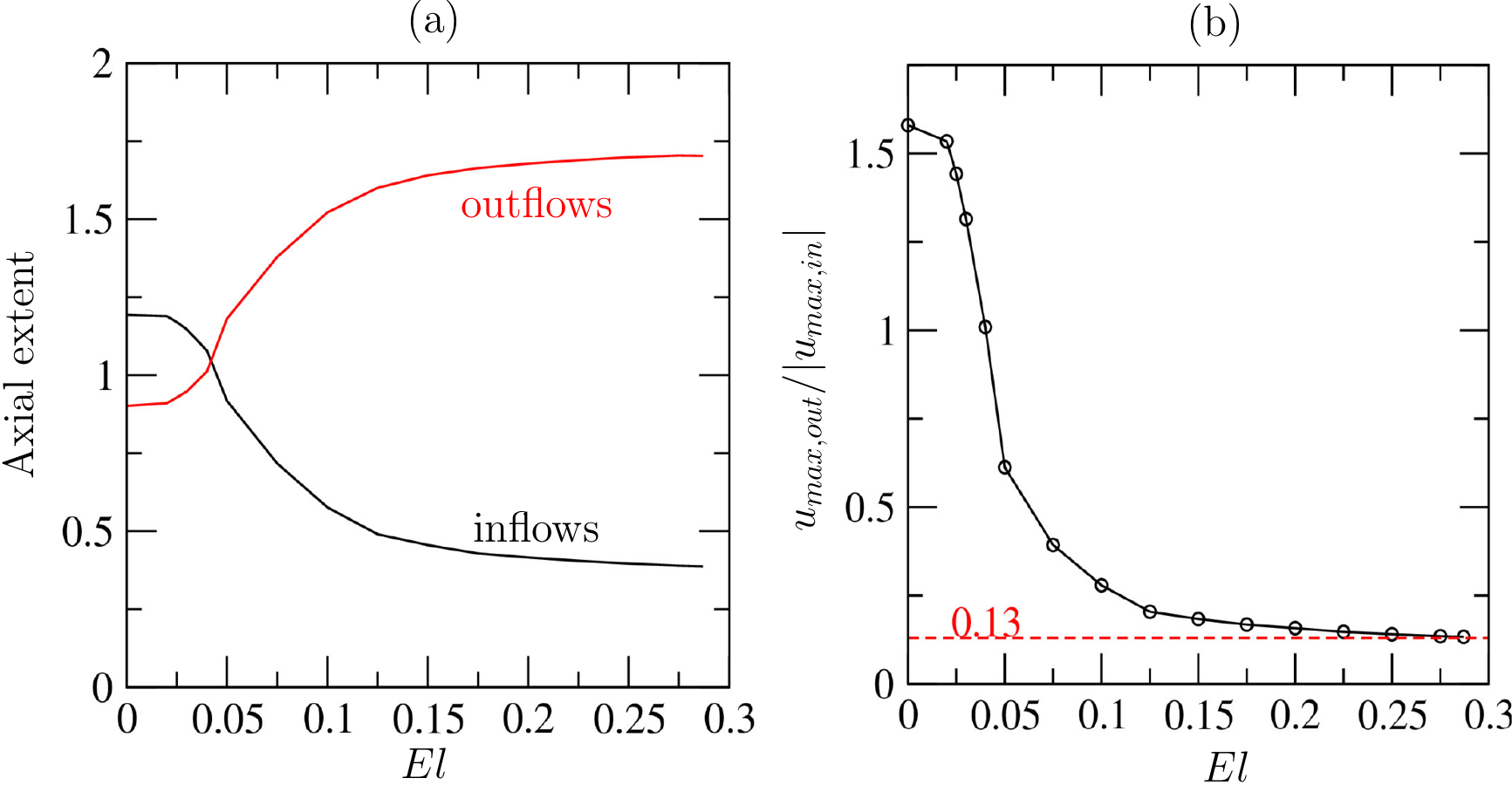}
  \end{center}
  \caption{ (Color online)  Quantification of the changes in the structure and strength of the vortices as $El$ increases. Panel $(a)$ shows the variation in the axial extent of inflows and outflows at the midgap in units of the gap-width, $d$, as the fluid becomes more elastic. Panel $(b)$ displays the ratio between the maximal values of $u$ in the outflows and inflows as a function of $El$. The dashed line indicates the approximate value this ratio seems to approach asymptotically as $El$ increases.}
  \label{fig:in_our_props} 
\end{figure}

\noindent A remarkable feature of this structural transition is the fact that the changes are most pronounced in the low $El$ regime ($El < 0.12$). This observation is quantitatively confirmed by the axial profiles of $u$ at the mid-gap shown in figure~\ref{fig:prof_rad_vel}.  
Profiles corresponding to $El < 0.12$ (shown in the panel $(a)$) differ markedly and clearly reflect strong changes in both the magnitude of $u$ (particularly in the outflows) and the axial extent of inflows and outflows. However, for $El \geq 0.12$ (see panel $(b)$), the differences among profiles are small and mainly occur in the magnitude of $u$, which keeps slightly decreasing (increasing) in the outflows (inflows) with increasing $El$. Further quantitative evidence of this behaviour is given in figure~\ref{fig:in_our_props}. The panel $(a)$ in this figure shows the dependence of the axial extent of the inflow and outflow regions at the midgap with increasing $El$. It is apparent that the largest variation in the extent of these regions (which are naturally inversely proportional) occur within the low $El$ regime, for $0.05 < El < 0.1$. Likewise, the sharpest change in the ratio between the maximal velocity of outflows and inflows (shown in the  panel $(b)$) also happens at very low $El$ values ($El < 0.05$), where the strength of the outflows decays strongly. These observations clearly evidence that $F_e$ exerts a surprisingly strong influence on the flow structure even in weakly elastic fluids. Another interesting feature revealed by the two panels of figure~\ref{fig:in_our_props} is that the flow characteristics appear to exhibit asymptotic behavior. As $El$ increases, the sizes of the outflows and inflows approach values close to $L_{out}/d \approx 1.7$ and $L_{in}/d \approx 0.38$, respectively, whereas the ratio between the maximal velocities in outflows and inflows appears to level off at approximately $0.13$. This observation is also corroborated by the velocity profiles, which seem to be gradually converging with increasing $El$ (see panel $(b)$ in figure~\ref{fig:prof_rad_vel}).\\

\begin{figure}
\centering
      \includegraphics[width=0.8\linewidth]{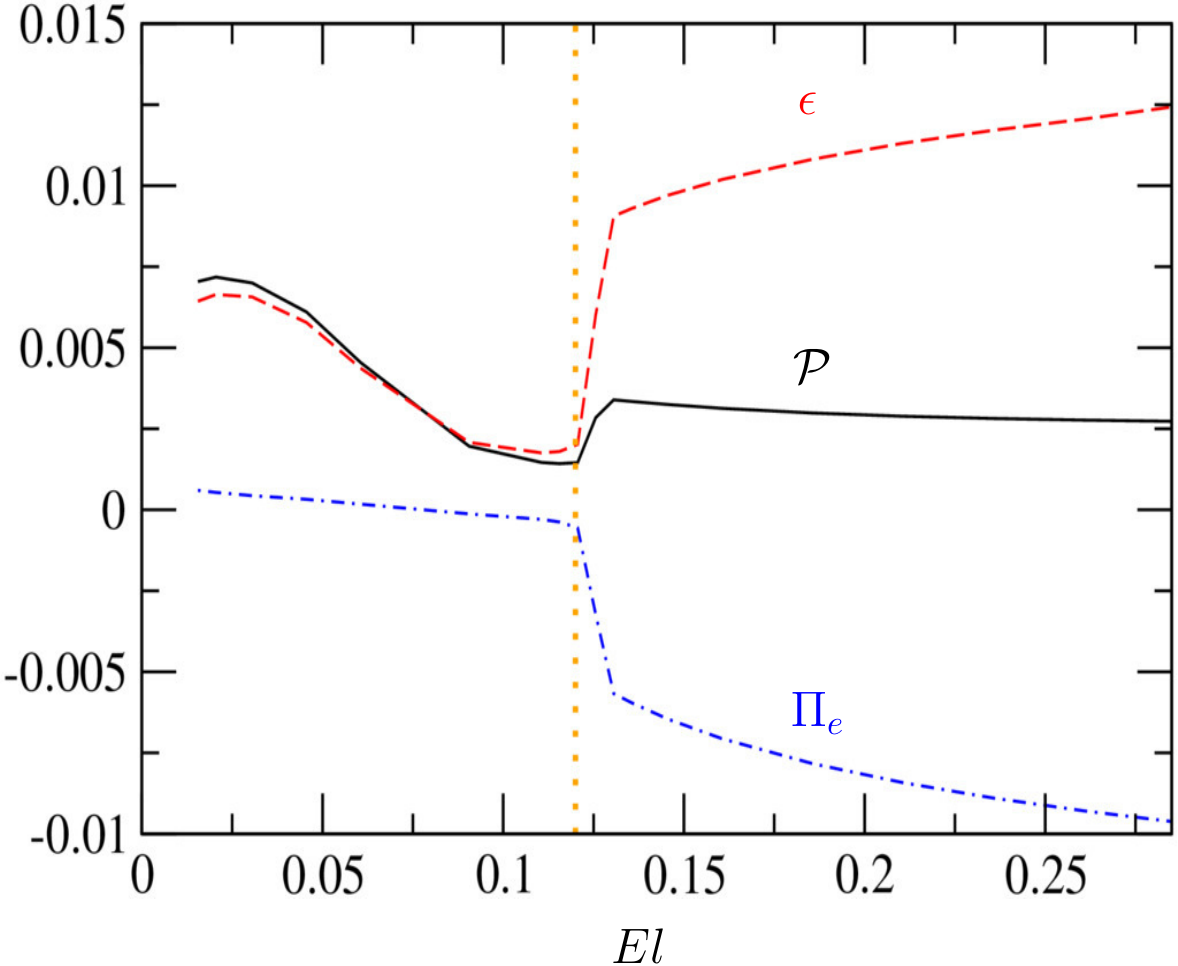} 
  \caption{ (Color online) Variation of the integral energy budgets with increasing $El$ before the spatio-temporal dynamics sets in. $\mathcal{P}$ denotes the kinetic energy production by the inertial forces, $\epsilon$ stands for the viscous dissipation and $\Pi_e$ is the work done by the elastic stresses. The dotted (orange) vertical line indicates the $El$ number at which the transition between the low and high $El$ regimes takes place. Hereafter, these regimes are denoted as centrifugally and elasticity dominated regimes, respectively.}
  \label{fig:eneg_balance}
\end{figure}

\noindent An important distinction between the low and high $El$ regimes (i.e. $El < 0.12$ and $El \geq 0.12$) is the magnitude of $u$ in the inflows. As seen in the panel $(a)$ of figure~\ref{fig:prof_rad_vel}, the maximum velocity of the inflows in the low $El$ regime increases initially with increasing $El$, but eventually converges to a value close to $u/|u_{max}| = -0.55$. This value is substantially lower than those shown by the profiles in the high $El$ regime (see panel $(b)$ in figure~\ref{fig:prof_rad_vel}), where $u/|u_{max}|$ ranges from $-1$ at $El = 0.125$ (when it is maximal) to $\sim -0.78$ at $El > 0.25$ (when the profiles seem to converge). The transition between both regimes can be clearly identified in the space time plot of figure~\ref{fig:spacetime_El} $(a)$ as a sudden change in the color intensity that takes place at $El \sim 0.12$. The abrupt nature of this transition strongly suggests that it may be caused by a change in the physical mechanism associated with the instability of the primary flow. To test this hypothesis we examine the integral energy budgets. For viscoelastic flows, the energy balance reads~\citep{dallas2010strong,Dubief13},
\begin{equation}\label{eq:eneg_balance}
  \int_V \mathcal{P} dV - \int_V \epsilon dV -  \int_V \Pi_e dV  = 0,
\end{equation}
where $\mathcal{P}$ is the kinetic energy production, $\epsilon$ is the viscous dissipation rate and $\Pi_e$ denotes the work done by 
the elastic stresses. These quantities were calculated using the following expressions: 
\begin{align}
     & \mathcal{P} = -\overline{u'v'} \frac{\partial \overline{v}}{\partial r} + \overline{u'v'} \frac{\overline{v}}{r},\label{eq:prod}\\
     & \epsilon = \frac{2\beta}{Re} \overline{S':S'},\label{eq:diss}\\
     & \Pi_e = \frac{1-\beta}{Re} \overline{S':T'}. \label{eq:poly_work}
\end{align}
Here, the overline denotes axially averaged quantities,  $S' = (\nabla \mathbf{v}' + \nabla \mathbf{v}'^T)/2$ is the rate of strain tensor and the prime symbol indicates deviations of the velocity or polymer stress tensor from their axially averaged values ($\overline{\mathbf{v}}=(0,\overline{v},0)$ and $\overline{\mathbf{T}}$). It must be clarified that, although this equation was derived in the context of turbulent flow, it also applies to steady and axisymmetric vortex flow (the derivation of the equation for this particular case is given in the appendix~\ref{appA}). The first and second integrals in equation~\ref{eq:eneg_balance} are always positive, meaning that they act as source and sink terms of the energy balance, respectively (note that there is minus sign in front of the second integral). The sign of the third integral can be positive or negative. If it is positive, this term has a negative contribution to the balance and thus polymers act to dissipate the fluid's kinetic energy. By contrast, if it is negative, polymers act as an energy source. The variation of the values yielded by these integrals with increasing $El$ is shown in figure~\ref{fig:eneg_balance}. As expected, the behaviour of the polymers changes drastically at $El \sim 0.12$, consistent with the transition between the low and high elasticity regimes. 
At low $El$ values, polymers play a dissipative role, helping the viscous forces to damp the centrifugally induced vortices. However, at $El \sim 0.1$, the work done by the elastic stresses changes sign and the net contribution of the polymers 
to the energy balance becomes positive, indicating that they inject energy into the flow through the elastic stresses. The amount of energy that the polymers supply to the system is initially very small (for $0.1 \leq El < 0.12$) but  increases  suddenly when $El \sim 0.12$. After this transition occurs, the energetic contribution of the polymers becomes the dominant energy source and its magnitude continues increasing with increasing $El$. In  contrast, the energy production 
due to inertial mechanisms remains small and decreases very gradually as $El$ increases. 
From this analysis, it is clear that the nature of the mechanism driving the instability indeed changes from being centrifugal ($El < 0.12$) to being elastic ($El \geq 0.12$), a characteristic that sets a clear distinction between the two regimes investigated so far.\\

\begin{figure}
\centering
      \includegraphics[width=\linewidth]{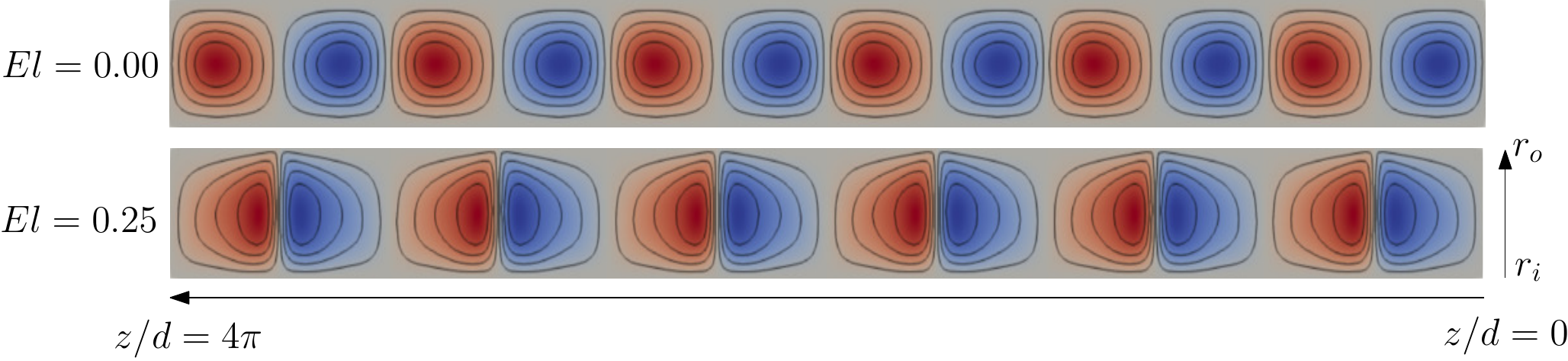} 
  \caption{ (Color online) Comparison between the flow streamlines $\psi$ in the centrifugally (top) and elastically (bottom) dominated regimes. Positive (negative) values are represented as red (blue), whereas the zero value is shown as gray. The color scale is based on the maximum and minimum values of each case and hence differs between both panels. There are 4 positive and 4 negative contours evenly distributed across the full range of values in each case: 1. $El = 0.00$, $\psi$ in $[-0.660,0.660]$; 2. $El = 0.25$, $\psi$ in $[-0.295,0.295]$. The system is shown rotated by 90 degrees in the counterclockwise direction.}
  \label{fig:st_comp}
\end{figure}

\noindent Finally, to facilitate comparison between the flow structure in the high $El$ regime and other elastically induced stationary patterns previously reported, the streamlines of the flow $\psi$ at $El = 0.25$ are shown in the bottom panel of figure~\ref{fig:st_comp}. These are naturally very different from those in the Newtonian case (also shown for comparison in the top panel) and reflect again the structural changes just discussed. As seen, unlike the Newtonian case, where the vortices are nearly equidistant, in the elastically dominated regime they appear arranged in pairs, with their cores being located very close to one another. We note that this type of structure has been previously reported in the literature and it is usually known as diwhirls (DW)~\citep{GroiStei97,LaBru01,ThoSurKho06,thomas_khomami_sureshkumar_2009}, due to its similarity with the shape of a magnetic dipole. However, there are a couple of important differences between the structures described in previous works and the one presented here. A characteristic shared by all previous studies is that DW appear after a hysteretic transition, when $Re$ is decreased starting from a flow state driven by an elastic instability. In fact, it is often stated in the literature that flow deceleration is a necessary condition to observe these structures~\citep{GroiStei97,LaBru01,ThoSurKho06,LaCaGiBa20}. The present study shows that this is not the case and that at least in the regime investigated here these structures may also appear for a fixed $Re$ if the elasticiy of the working fluid is sufficiently large so that the elastic instability mechanism replaces the centrifugal mechanism. In the low $Re$ regimes where most previous studies were conducted, DW appear localized in the axial direction, i.e. there are regions where the flow is laminar interspersed between distinct DW. When the distance between DW becomes less than $5d$, they approach each other and coalesce. This characteristic is so far absent in the present simulation. Despite the distance between DW is substantially lower than $5d$, they remain stationary and form a  pattern of equally spaced structures along the axial direction. A possible reason for such difference will be discussed later in section~\ref{sec:momentum}. We would like to finally note that similar arrangements of DW, yet not stationary, have also been reported in the literature, which were dubbed oscillatory strips~\citep{GroStei96,ThoSurKho06,thomas_khomami_sureshkumar_2009}. 

\subsection{Onset of spatio-temporal dynamics}\label{sec:complex}

\begin{figure}
      \includegraphics[width=\linewidth]{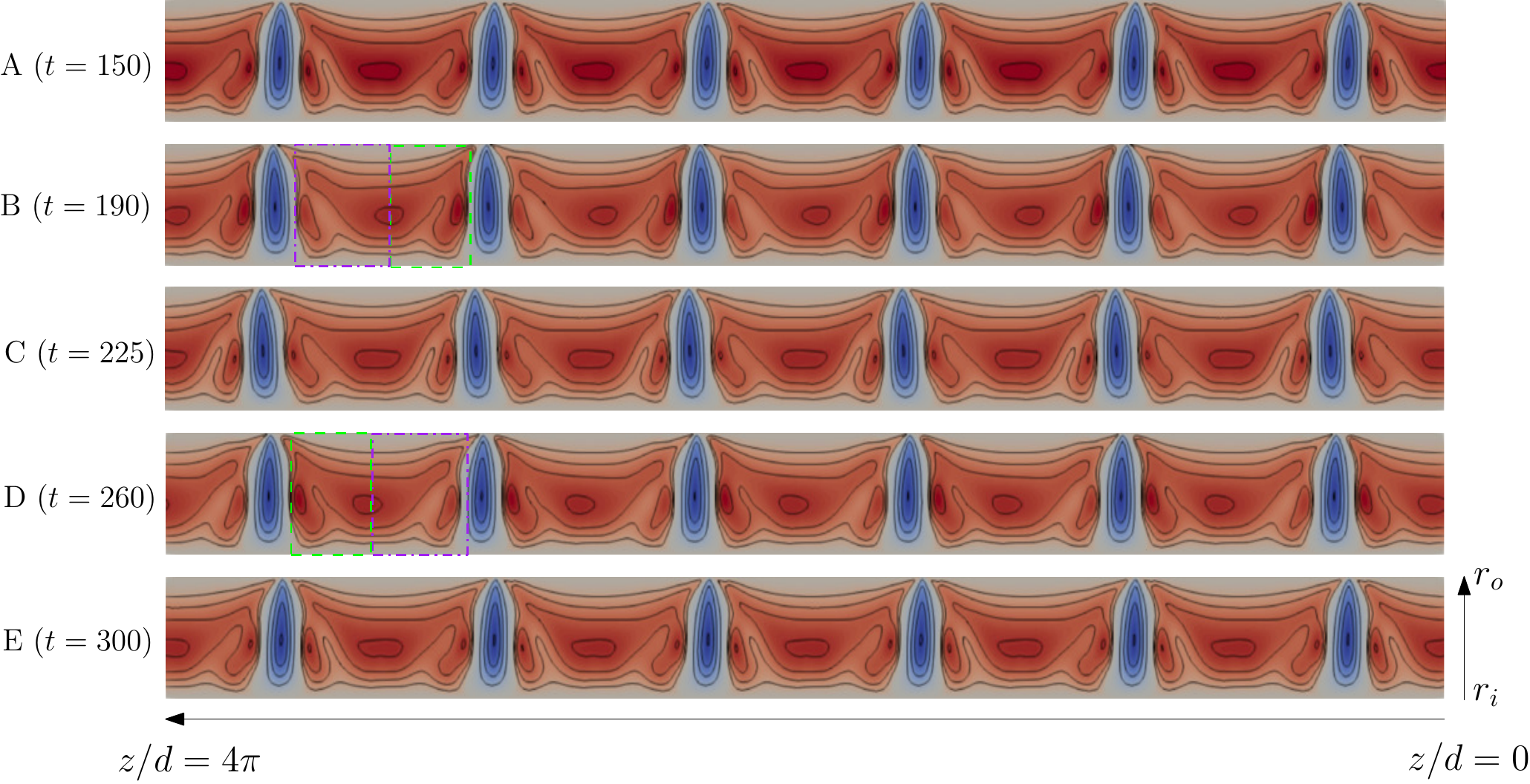} 
      \caption{ (Color online) Color maps of $u$ illustrating the axial oscillation of the vortex pairs that emerges from the Hopf bifurcation. Letters from A to E are used to show the correspondence between each panel and the time series shown in figure~\ref{fig:periodic}.  Positive (negative) velocity is represented as red (blue), whereas the zero velocity is shown as gray. There are $4$ positive and $4$ negative contours evenly distributed in $u \in [-0.0780,0.0095]$. The system is shown rotated by $90$ degrees in the counterclockwise direction. Note that the flow patterns shown in the panels $A$ and $E$ are identical, whereas those shown in the panels B and D only differ in that their outflows are antisymmetric.}
      \label{fig:axial_shift}
\end{figure}

\begin{figure}\setlength{\piclen}{0.45\linewidth}
  \begin{center}
    \begin{tabular}{c c}
      $(a)$  & $(b)$\\
      \includegraphics[width=\piclen]{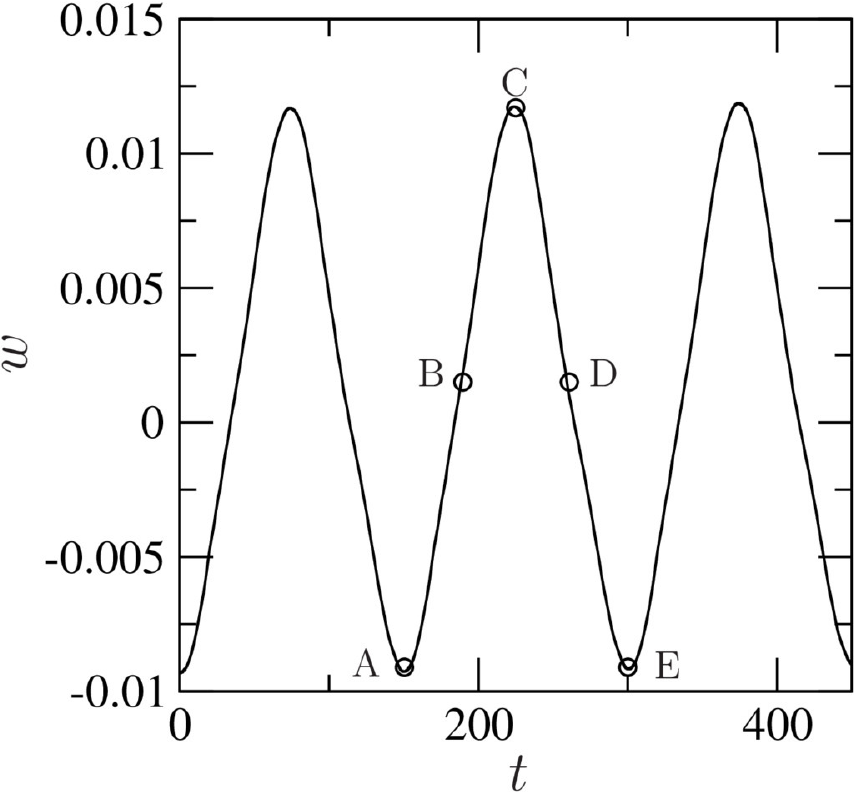} &
      \includegraphics[width=\piclen]{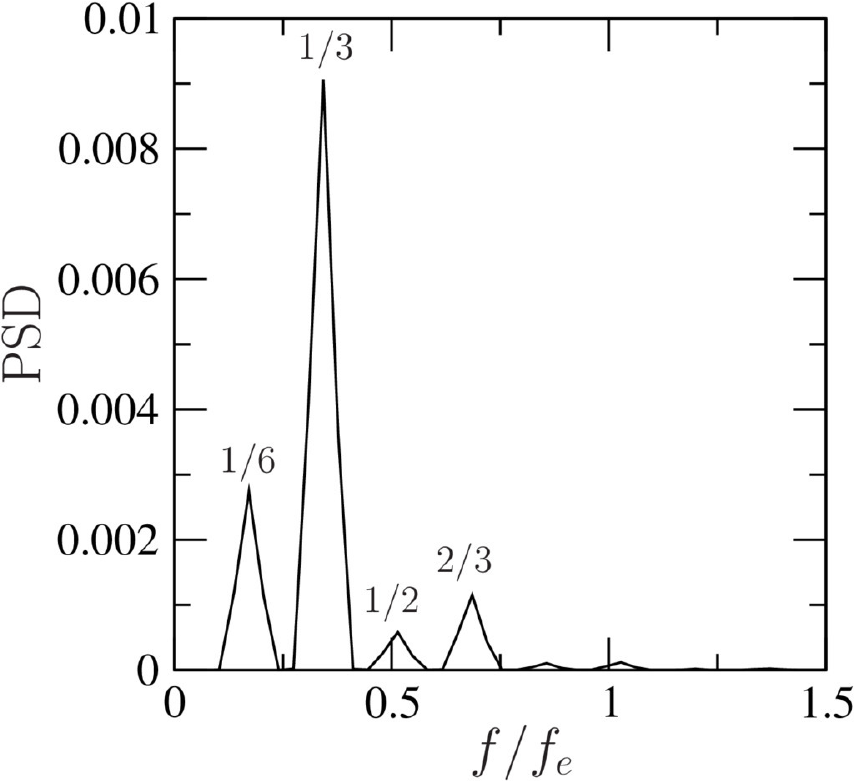}\\
    \end{tabular}
  \end{center}
  \caption{Periodic motion arising from from the instability of the stationary pattern of diwhirls. $(a)$ Time series of the axial 
  velocity $w$ for $El = 0.3$ calculated at $r/d = 4.15$ and $z/d = \pi$ . The circles indicate time instants for which colormaps of 
  $u$ are presented in the figure~\ref{fig:axial_shift}. $(b)$ Power spectral density (PSD). The frequency is normalized with the 
  elastic frequency, $f_e$.}
  \label{fig:periodic} 
\end{figure}

\noindent The stationary pattern of DW loses its stability at a Hopf bifurcation which takes place at $El \sim 0.29$ leading to an axial oscillation of the vortices. This is illustrated in figure~\ref{fig:axial_shift} through color maps of $u$ taken at five equally spaced time instants within a period (shown as circles in the panel $(a)$ of figure~\ref{fig:periodic}).  The displacement of the vortices is relatively small, yet clearly discernible in the figure by looking at the axial position of the inflows. It should be recalled that the system is shown rotated by 90 degrees counterclockwise, and so the upwards (downwards) motion of the inflows corresponds to leftwards (rightwards) motion in these figures. Starting from the state where the inflows are at their lowest axial positions (panel $A$), it is observed that the inflows move first axially upwards (panel $B$), reach the position of maximum displacement (panel $C$) and subsequently move downwards (panel $D$), returning eventually to the initial state (panel $E$, which is identical to $A$).  A notable difference with respect to the stationary case is the breakdown of the symmetry in the outflows. When the vortices move axially upwards, the lower half of the outflow  (see region enclosed by the (green) dashed rectangle in the panel B) remains similar to that in the stationary case (see bottom panel in figure~\ref{fig:cmap_radial_velocity}), however the upper half (marked by the (purple) dash-dotted rectangle in the panel B) notably changes due to an increase in the maximal velocity next to the inflow. The opposite is observed when the vortices move downwards. The upper half (shown as a (green) dashed rectangle in the panel D) of the outflow remains as in the  stationary case, whereas the lower half ((purple) dash-dotted rectangle in the panel D) takes a similar form to that of the upper half during the upward motion. As a consequence, flow states moving axially upwards and downwards, where the vortices are located at the same axial positions, exhibit antisymmetric outflows (that is the case, for example, for the states B and D shown in the figure).\\

\noindent The frequency of the oscillation was determined by applying the Fast Fourier transform to a time series of the axial velocity $w$ obtained at a radial location close to the outer cylinder (figure~\ref{fig:periodic} $(a)$). The power spectral density is shown in figure~\ref{fig:periodic} $(b)$, where the frequency is normalized with the elastic frequency, $f_e$. Following~\cite{LaCaGiBa20}, $f_e$ was calculated as $f_e = 2c_e/k_{avg}$, where $c_e$ denotes the wave celerity, $c_e = \sqrt{\nu/\lambda}$, and $k_{avg}$ is the average spatial wavelength of the vortex flow pattern. The spectrum shows a pronounced peak at $f_e/3$, which clearly indicates that this is the dominant frequency of the oscillation. Other peaks with smaller amplitudes are also observed. However, they correspond in all cases to other sub-harmonics of $f_e$ and are therefore commensurate with the dominant frequency. It should be noted that $f_e/3$ is the dominant frequency for the particular case where the flow pattern has $6$ vortex pairs. 
For other flow patterns with different number of vortex pairs, the oscillation is characterized by other subharmonics of $f_e$. The fact that frequencies appear in the spectrum as subharmonics of $f_e$ is in full agreement with the experimental observations~\citep{LaCaGiBa20} and reflects once again the elastic nature of the instabilities taking place at these $El$ values. \\

\begin{figure}
      \includegraphics[width=\linewidth]{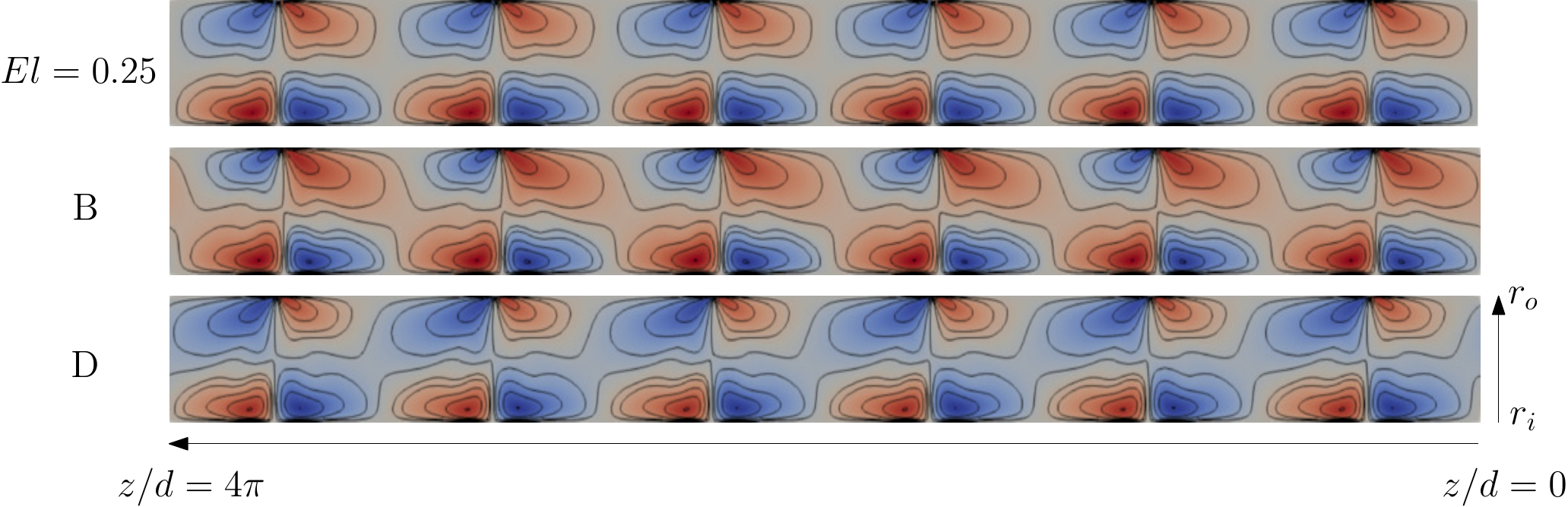} 
  \caption{ (Color online) Color maps of $w$ illustrating the standing wave causing the axial oscillation of the flow pattern. The top panel shows $w$ for a stationary state at $El = 0.25$, whereas the middle and bottom panels correspond to the states B and D indicated in the figure~\ref{fig:periodic}.  Positive (negative) values of $w$ are shown as red (blue), whereas that for $w=0$ is shown as gray. From top to bottom, there are $4$ positive and $4$ negative contours evenly distributed in $w \in [-0.030,0.030]$, $[-0.026,0.030]$ and $[-0.030,0.026]$, respectively. Two additional contours at $w = \pm0.003$ are also added to help visualize the axial waves. The system is shown rotated by $90$ degrees in the counterclockwise direction.}
  \label{fig:axial_vel}
\end{figure}

\noindent From a mechanistic perspective, the periodic up and down motion of the vortices is just a consequence of the physics described in the previous section. The distance between the centres of the vortices on either side of the inflows keeps decreasing (albeit very gradually) as $El$ increases, leaving a gap between DW where the radial velocity is increasingly weak. The axial velocity, whose role before the instability onset is simply to transport the fluid vertically near the cylinders (see top panel in figure~\ref{fig:axial_vel}), eventually penetrates into these intermediate regions, connecting adjacent vortices to each other in the outflow region and giving rise  to an axial wave. This wave propagates first axially upwards (middle panel in figure~\ref{fig:axial_vel}) and subsequently reflects back and travels axially downwards (bottom panel in figure~\ref{fig:axial_vel}), thereby creating a standing wave. The interaction between standing wave and vortices lead to the axial oscillation of the latter illustrated above.\\

\subsection{Vortex merging and splitting dynamics}\label{sec:merging_splitting}

\begin{figure}
    \includegraphics[width=\linewidth]{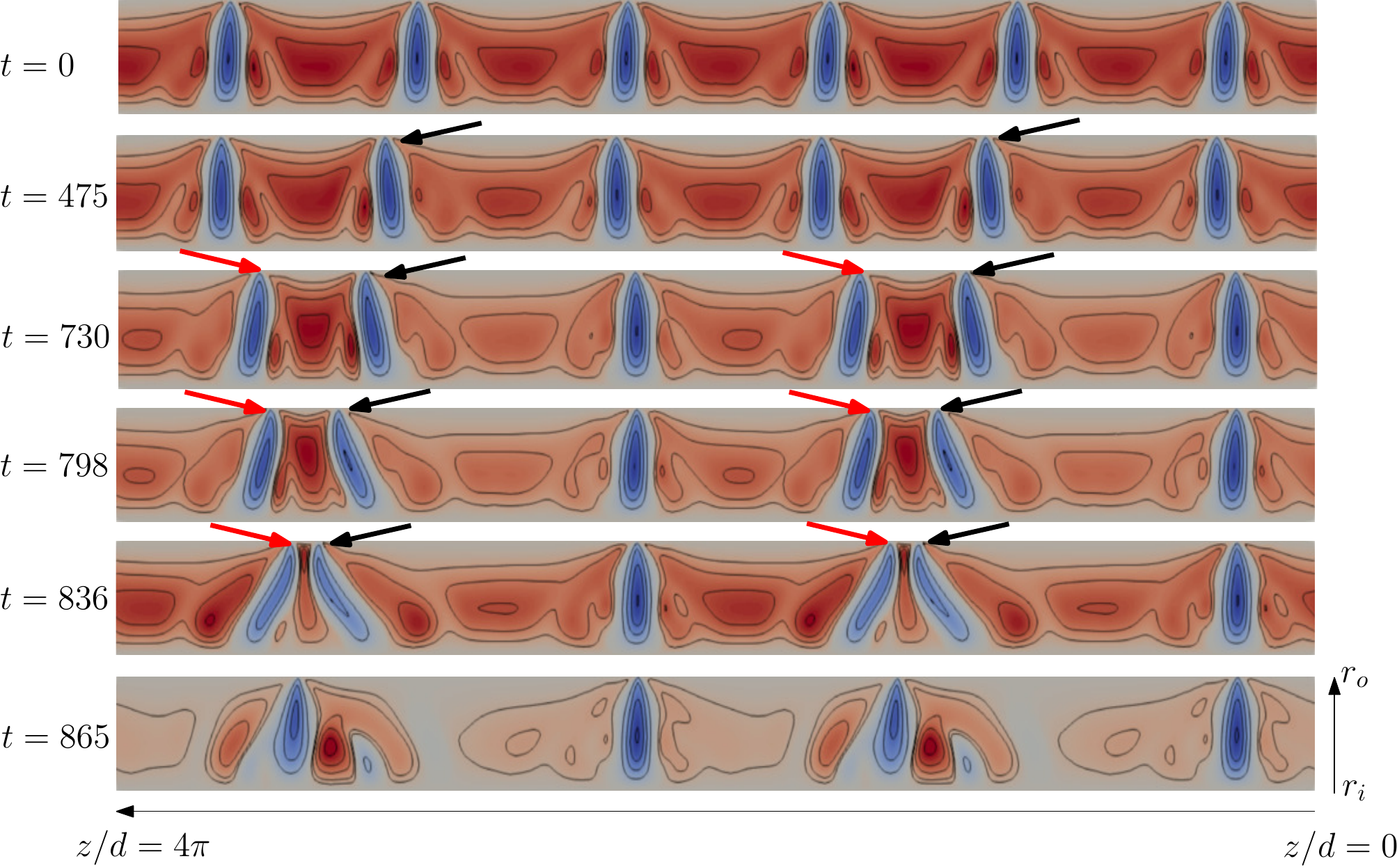} 
    \caption{ (Color online) Color maps of $u$ illustrating the simultaneous occurrence of two merging events at the beginning of the VMS regime. They correspond to a simulation performed at constant $El = 0.32$, whose space-time diagram was shown in the panel (c) of the figure~\ref{fig:spacetime_El}. The dashed (green) rectangle in the latter figure indicates the time window that is illustrated in the current figure. Positive (negative) values of $u$ are shown as red (blue), whereas $u=0$ is shown as gray. In each panel there are $4$ positive and $4$ negative contours evenly distributed in $u \in [-0.071,0.013]$. The system is shown rotated by $90$ degrees in the counterclockwise direction.}
  \label{fig:merging_event}
\end{figure}

With a further increase in $El$, the dynamics of the different DW decouple and these begin to travel independently along the axial direction. A merging event happens when two DW move towards each other and eventually coalesce into a single entity. This phenomenon is illustrated in figure~\ref{fig:merging_event}, which shows the variation of the radial velocity over the initial stage of the VMS regime in the simulation performed at constant $El=0.32$. Two merging events occur simultaneously in the time window shown (corresponding to the (green) dashed rectangle in the panel (c) of the figure~\ref{fig:spacetime_El}). In the initial phase of the merging process (up to $t \approx 500$), the second and fifth DW starting from the top (indicated by (black) leftwards arrows) leave their axial locations and begin to move upwards, i.e. leftwards in the rotated figure. This motion becomes apparent by comparing the positions of the inflows between the first ($t=0$) and second ($t=475$) panels. The inflows of the second and fifth DW have notably moved towards the top of the system by $t = 475$ and differ from the others in that they are tilted slightly upwards (a distinctive feature of the DW which are moving upwards). The inflows of the other DW on the other hand remain at their axial positions and only exhibit small oscillations caused by the instability discussed in the section above. As the cores of the DW travelling upwards approach the cores of the adjacent DW, they experience an attractive interaction which results in the first and fourth DW (indicated by (red) rightwards arrows) travelling axially downwards, i.e. rightwards in the rotated figure. Note that, as opposed to the DW moving upwards, the inflows of the DW moving downwards are tilted downwards. The attractive interaction between DW becomes stronger as their cores get close to each other, leading to a rapid increase of their travelling speeds. This characteristic results in the typical parabolic shape exhibited by the merging events in the space-time plot shown in figure~\ref{fig:spacetime_El}. Another feature that is clearly illustrated in the figure~\ref{fig:merging_event} is the increase in the tilting angle of the inflows as the DW approach one another. This angle is initially very small (see e.g. panel for $t = 730$) but increases rapidly as the mutual interaction between DW becomes stronger, reaching a value of approximately $30$ degrees with respect to the radial axis by the time when the merging of the DW takes place (see panel for $t = 836$). \\

\noindent After the merging events are completed, a flow pattern characterized by four DW emerges (see panel for $t = 865$), which retains a discrete translational symmetry along the $z$-direction (i.e. the flow pattern remains invariant if it is shifted by $2\pi$ in the axial direction). As can be seen from the panel (c) of figure \ref{fig:spacetime_El}, this flow pattern undergoes subsequent merging events, which occur again simultaneously, when $t$ is between $865$ and $2320$. After this second pair of merging events, the axial symmetry of the flow is fully broken (not shown) and the dynamics of the distinct DW are fully decoupled. It is only after the latter happens that the sequence of merging and splitting events begins and the number of DW in the system may either grow or decay. This initial cascade of merging events leading to a complete symmetry breaking is always observed in the simulations at the beginning of the VMS regime and can thus be interpreted as a transitional stage where the coupling between DW is fully broken. It must be finally noted that the time scale associated with merging events is highly variable and depends crucially on the distance between the cores of the two DW undergoing the merging event. They may occur very slowly,  extending over nearly $1000$ advective time units, such as the merging process exemplified in the figure~\ref{fig:merging_event}, or very quickly,  within $50$ to $100$ advective time units, like the ones discussed in the next paragraph.\\

\begin{figure}
      \includegraphics[width=\linewidth]{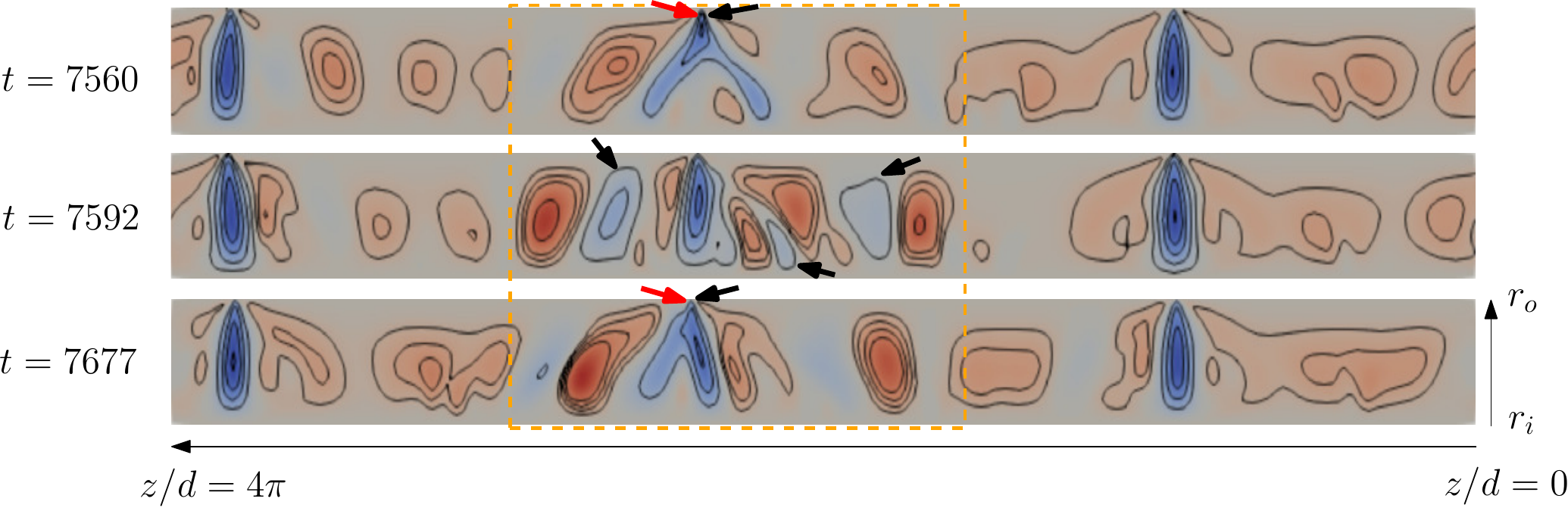} 
  \caption{ (Color online) Color maps of $u$ illustrating the flow patterns associated with the chaotic dynamics that appear interspersed between slow merging events and splitting events. The dynamics displayed in this figure corresponds to the simulation with constant $El=0.32$, and more specifically, to the time window shown as a solid line (brown) rectangle in the panel (c) of the figure~\ref{fig:spacetime_El}. Positive (negative) values of $u$ are shown as red (blue), whereas $u=0$ is shown as gray. There are $4$ positive and $4$ negative contours evenly distributed in $u \in [-0.073,0.013]$. The system is shown rotated by $90$ degrees in the counterclockwise direction.} 
  \label{fig:transient}
\end{figure}

\noindent A central aspect of the dynamics in the VMS regime is the emergence of transient chaotic motion interspersed between merging and splitting events. As seen in figure~\ref{fig:spacetime_El},  this chaotic dynamics appears localized in the vicinity of the region where a merging event has taken place and it is characterized by the repeated emergence of closely spaced, weak vortices which merge shortly after they form, creating a quick succession of merging and splitting events. The characteristic cycle of this transient dynamics is exemplified in the figure~\ref{fig:transient}, which shows the evolution of the flow pattern in a narrow temporal window of the simulation performed at $El = 0.32$ (see the solid line (brown) rectangle in the panel (c) of the figure~\ref{fig:spacetime_El}). In this example, the chaotic dynamics occurs in the central part of the system, framed by a dashed (orange) rectangle in the figure, and has little effect on the DW located outside this region. When a merging event is accomplished (see upper panel), the energy released by the DW which is eliminated is transferred to an irregular wavy motion. This wave transports the energy axially upwards and downwards from the location where the merging took place and results in the formation of new vortices (the inflows of the newly created vortex pairs are indicated with arrows in the intermediate panel). The amount of time it takes from the end of the merging event until the new vortices are fully formed normally ranges between $20$ and $30$ advective time units. The new vortices are however just a small distance apart from each other, so that they undergo a strong attractive interaction and quickly merge (see lower panel). The energy released after the new merging events is again redistributed and the process just described starts over again.  The time span between consecutive merging events is of the order of $100$ advective time units, whereas the entire time period over which the chaotic dynamics typically extends ranges from $1000$ to $2000$ advective time units.\\

\begin{figure}
      \includegraphics[width=\linewidth]{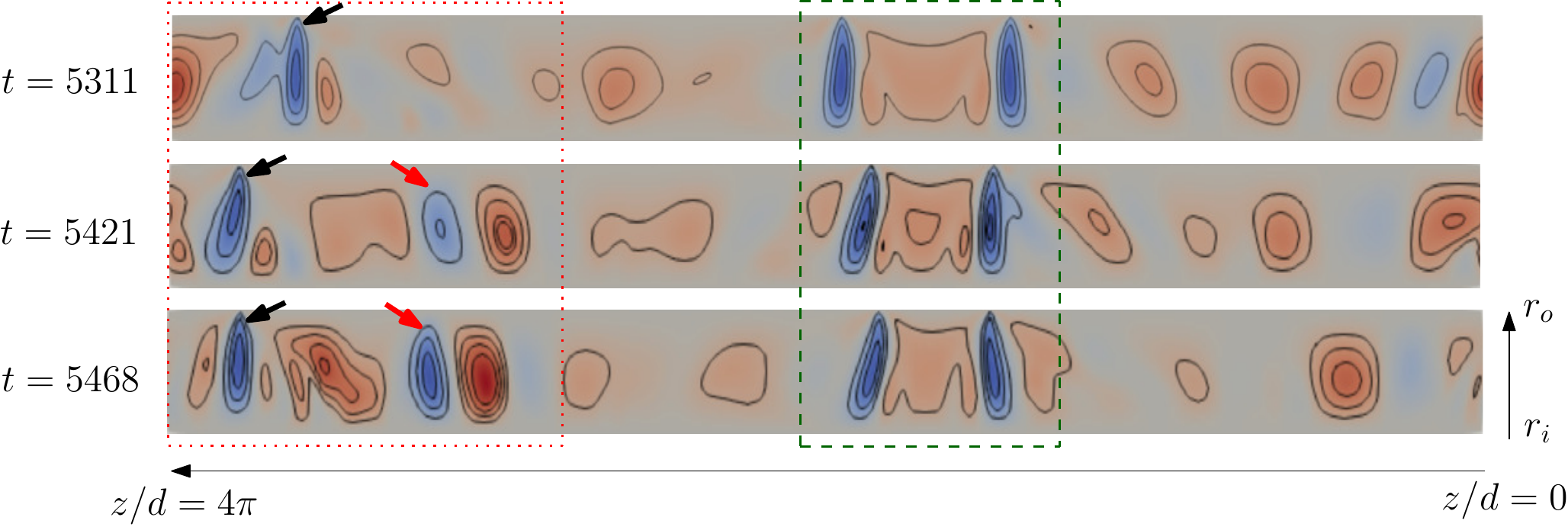} 
  \caption{ (Color online) Color maps of $u$ illustrating the occurrence of a splitting event. The example shown corresponds to the simulation with constant $El=0.32$ and it is indicated as a dash-dotted (purple) rectangle in the panel (c) of the figure~\ref{fig:spacetime_El}. Positive (negative) values of $u$ are shown as red (blue), whereas $u=0$ is shown as gray. There are $4$ positive and $4$ negative contours evenly distributed in $u \in [-0.077,0.015]$. The system is shown rotated by $90$ degrees in the counterclockwise direction.}
  \label{fig:splitting}
\end{figure}

\noindent The main splitting events (i.e. events where a new, strong and persistent DW is created) are normally preceded by chaotic motion and take place when the distance between newly created DW, as well as the distance between these and the other DW in the system, become sufficiently large so that their mutual interaction is weak. An example of a splitting event taking place between $t = 5300$ and $t = 5470$ in the simulation performed at $El=0.32$ (see the dash-dotted (purple) rectangle in the panel (c) of the figure~\ref{fig:spacetime_El}) is shown in figure~\ref{fig:splitting}. The splitting event occurs in the region marked by the (red) dotted rectangle. The DW that appears within this region in the upper panel of the figure (indicated by a leftwards arrow) is the result of a merging event which has just completed. The other two DW in the figure (which are enclosed in a (green) dashed rectangle) are moving towards each other as a part of a merging process that will be completed after the splitting event takes place. Hence, the distance between the core of the topmost DW and those of the other two DW becomes increasingly large with time. This enables that when a new DW appears in the space left between them (see intermediate panel,  where the new DW is indicated by a rightwards arrow), it can be sufficiently far apart from its neighbors to avoid a strong attractive interaction. As a result, the new DW does not undergo any merging events shortly (in contrast to what happens during the transient chaotic motion) and its strength increases until it becomes comparable to that of the other DW in the system (see lower panel). The new DW is connected to the topmost DW through the outflows, which is is an indication that these DW will eventually undergo a merging event. However, such event happens at $t \sim 5930$ (see panel (c) in the figure~\ref{fig:spacetime_El}), nearly $500$ advective time units after the new DW first appeared. In general, the DW created after primary splitting events are found to persist for several hundred advective time units, as opposed to the vortices created during the chaotic dynamics which never persist longer than a few tens of advective time units.\\

\begin{figure}
      \includegraphics[width=\linewidth]{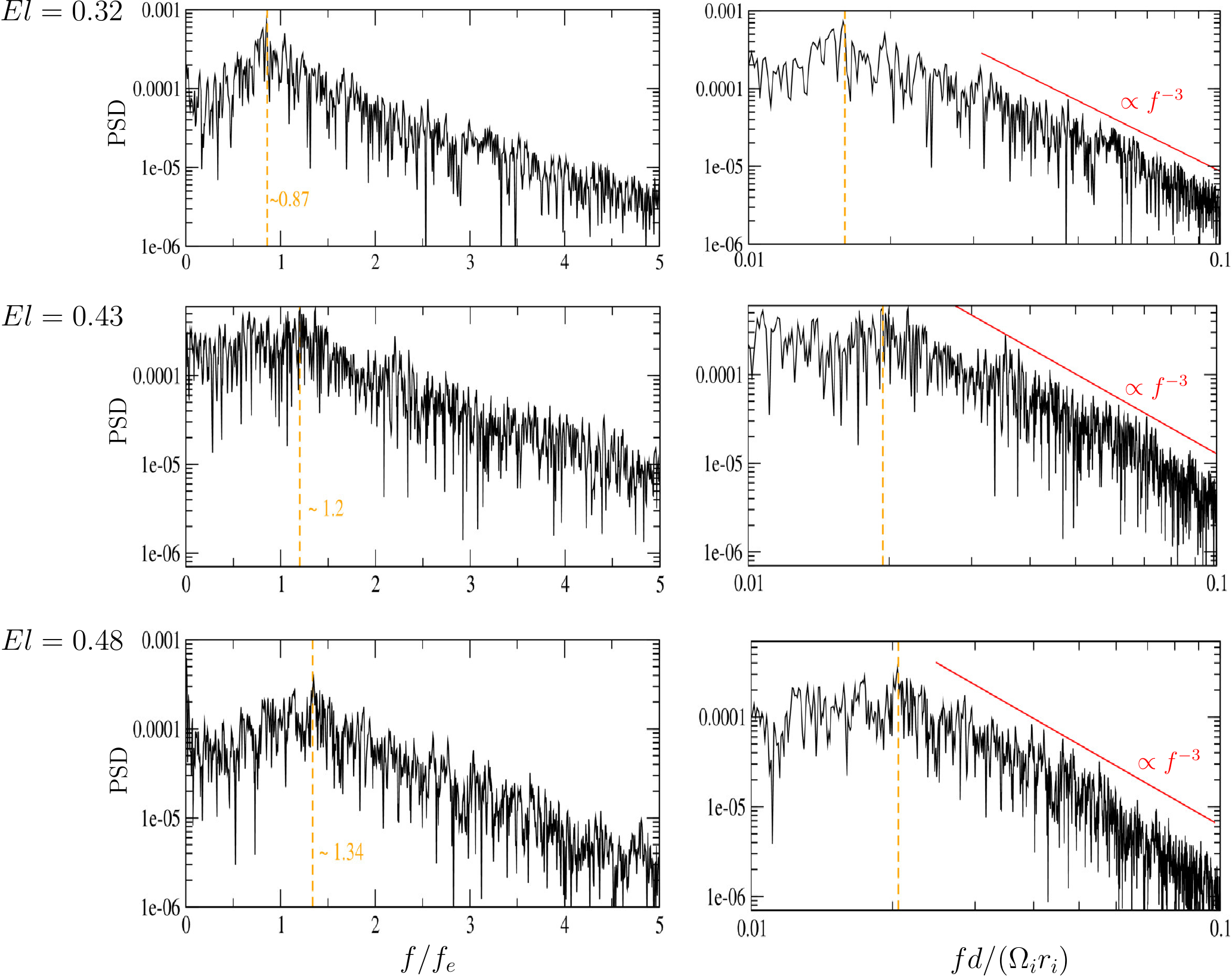} 
  \caption{ (Color online)  Power spectral density (PSD) obtained from simulations in the VMS regime at three distinct values of $El$. The left panels show the spectra in a linear-log plot, with the frequency normalized with the elastic frequency, $f_e$, whereas the right panels show the spectra in log-log scale and the frequency is non-dimensionalized with the inverse of the advective time.} 
  \label{fig:freq_VMST}
\end{figure}

\noindent The dynamics illustrated by these examples repeats successively with time, creating a chaotic regime characterized by continuous changes in the number of DW (the regime that has been  dubbed VMS regime). Power spectral characterization of this flow regime is shown in the figure~\ref{fig:freq_VMST}. The spectra were computed by applying the fast Fourier transform to time series of the radial velocity at the mid-gap obtained from simulations where $El$ was held constant. The power spectral density increases at the lowest frequencies until it reaches a maximum (indicated as a dashed (orange) line in the figure). For the range of $El$ values investigated, the frequency at which the maximum takes place is close to $f_e$ and increases with increasing $El$, from $0.87 f_e$ at $El = 0.32$ to $1.36 f_e$ at  $El = 0.5$. After the peak, the power spectral density decreases monotonically and exhibits to a good approximation a power law decay range with a decay rate of $-3$ (the best fit to the data yields exponents ranging between $-2.82$ and $-3.33$). This exponent is in agreement with the universal spectral decay rate that was theoretically predicted for elasto-inertial turbulence~\citep{Foux03}, which has been recently verified in experiments~\citep{Ya21}, and thereby identifies the VMS regime as a class of elasto-inertial turbulent states. It must also be remarked the similarity of the power spectra obtained at different $El$ values. This observation suggests that flows in the VMS regime are not significantly affected by changes in $El$.\\

\subsection{Influence of the domain size and other computational parameters}\label{sec:domain_size_and_others}
\begin{figure}
  \includegraphics[width=\linewidth]{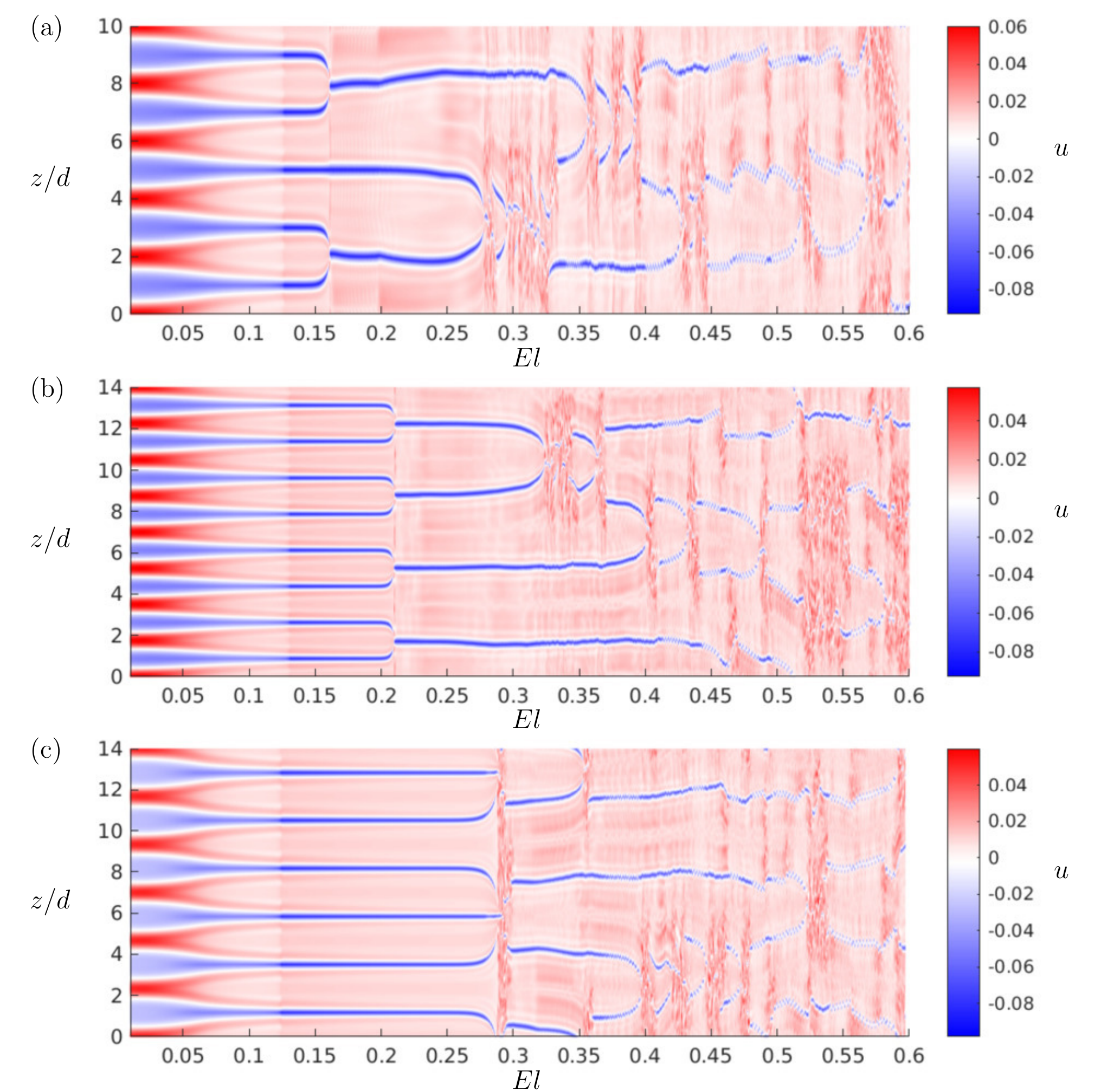} 
  \caption{ (Color online) Space-time diagrams showing the magnitude of $u$ at the mid-gap along the $z$ direction in simulations where the axial length of the system and/or the number of vortex pairs of the initial condition were varied with respect to the simulations shown in the figure~\ref{fig:spacetime_El}. The cases illustrated correspond to (a) $\Gamma = 10$ using an initial condition with $5$ vortex pairs, (b) $\Gamma = 14$ using  an initial condition with $8$ vortex pairs and (c) $\Gamma = 14$ using an initial condition with $6$ vortex pairs. As in the simulations presented in the figure~\ref{fig:spacetime_El}, $El$ has been slowly increased with time ($El = 10^{-3}t/Re$) and the latter has been replaced by its corresponding $El$ value in the horizontal axis of the space time plots. Red and blue areas indicate outflows and inflows respectively. Note that periodic boundary conditions are used in $z$.}
  \label{fig:domain_size}
\end{figure}

\noindent Since VMS events arise from interactions among DW, it is natural to wonder about the impact that changes in 
the length-to-gap aspect ratio $\Gamma$ (and consequently in the number of DW the system may contain) may have on the findings 
reported in the previous subsections. To investigate this aspect, a set of simulations where $\Gamma$ was varied between $9$ and $16$ 
has been conducted. The protocol followed in these simulations is the same as that described at the beginning of the 
section~\ref{sec:to_VMST}: they are started from a Newtonian Taylor vortex flow and the $El$ number is slowly increased 
with time at a uniform rate, $El = 10^{-3}t/Re$, where $Re$ was again fixed to $95$. The influence of the initial condition 
has also been examined. To this end, two or three simulations have been performed for each value of $\Gamma$ considered 
and these have been started from states having a different number of vortex pairs. A complex spatio-temporal dynamics 
consistent with the VMS regime has been observed in all cases. However, the $El$ threshold at which merging events first 
appear changes significantly depending on the domain size and the initial condition. This is illustrated in the 
figure~\ref{fig:domain_size}, which shows space time diagrams of $u$ at the mid-gap along the $z$ direction for simulations 
where (a) $\Gamma = 10$ and the initial condition has $5$ vortex pairs, (b) $\Gamma = 14$ and the initial condition has 
$8$ vortex pairs and (c) $\Gamma = 14$ and the initial condition has $6$ vortex pairs. Note that analogously to the 
figure~\ref{fig:spacetime_El} (a) time has been replaced by its corresponding $El$ value. As seen, the first merging events 
are accomplished at notably different values of the $El$ number in each case: $El \approx 0.16$ in (a),  
$El \approx 0.21$ in (b) and $El \approx 0.28$ in (c), which in turn differ from the onset of merging events 
reported in the section~\ref{sec:to_VMST} ($El \approx 0.32$ for $\Gamma = 12.56$, using a state with $6$ vortex pairs 
as initial condition). It is therefore evident that this feature is very sensitive to the domain size and the initial condition used 
in the simulations. The same is true for the onset of chaotic motion. For the $\Gamma$ values and initial conditions 
investigated, the first occurrence of a merging event has been found to range between $El \approx 0.15$ and 
$El \approx 0.32$, whereas chaotic motion has been first observed at $El$ values ranging from $0.25$ to $0.42$.

\noindent The characteristics of the transition towards the VMS regime are not significantly altered by changes in $\Gamma$ or the initial condition. As $El$ increases initially from the Newtonian limit, a centrifugally dominated regime is identified in all cases, where the axial extent of the inflows (outflows) gradually decreases (increases) with increasing $El$. This behaviour continues until the elastic instability sets in abruptly and the intensity of the inflows becomes much higher. This can be seen in the figure~\ref{fig:domain_size} as a sudden change in the color intensity that happens at $El \approx 0.125$. It is interesting to note that, despite the significant variation in the $El$ values at which the VMS events occur, the threshold of the elastic instability remains nearly unchanged with varying $\Gamma$ or initial condition (it is always found to occur at $El$ values ranging from $0.12$ to $0.13$). Another feature that is shared by all simulations regardless of the domain size and initial condition is the existence of an initial stage of merging events, with some of them taking place simultaneously, that precede the appearance of chaotic motion. A previously unreported event, where three DW merge simultaneously, has been observed in some simulations within this initial stage (see panel (c) of figure~\ref{fig:domain_size}). This particular type of merging event occurs when there is a group of $3$ equal DW in which the upper and lower DW move towards the central DW. The forces exerted by the upper and lower DW on the central DW are equal and act in opposite senses, so that the central DW does not move and eventually the three DW merge simultaneously. The main difference between the transition scenarios illustrated in the figure~\ref{fig:domain_size} and that described in the previous subsections is the absence of the regime characterized by the small axial oscillations of the flow pattern. This regime preceded the onset of the VMS regime in the simulation for $\Gamma = 12.56$. The same axial oscillations are observed in the figure~\ref{fig:domain_size} at large $El$ values. However, merging and splitting events occur prior to the emergence of the oscillations in these cases. This clearly indicates that the occurrence of these oscillations is not a necessary condition for the VMS dynamics to exist, but an additional dynamics that occurs at large $El$ values and may or not coexist with the VMS events.\\

\begin{figure}
      \includegraphics[width=\linewidth]{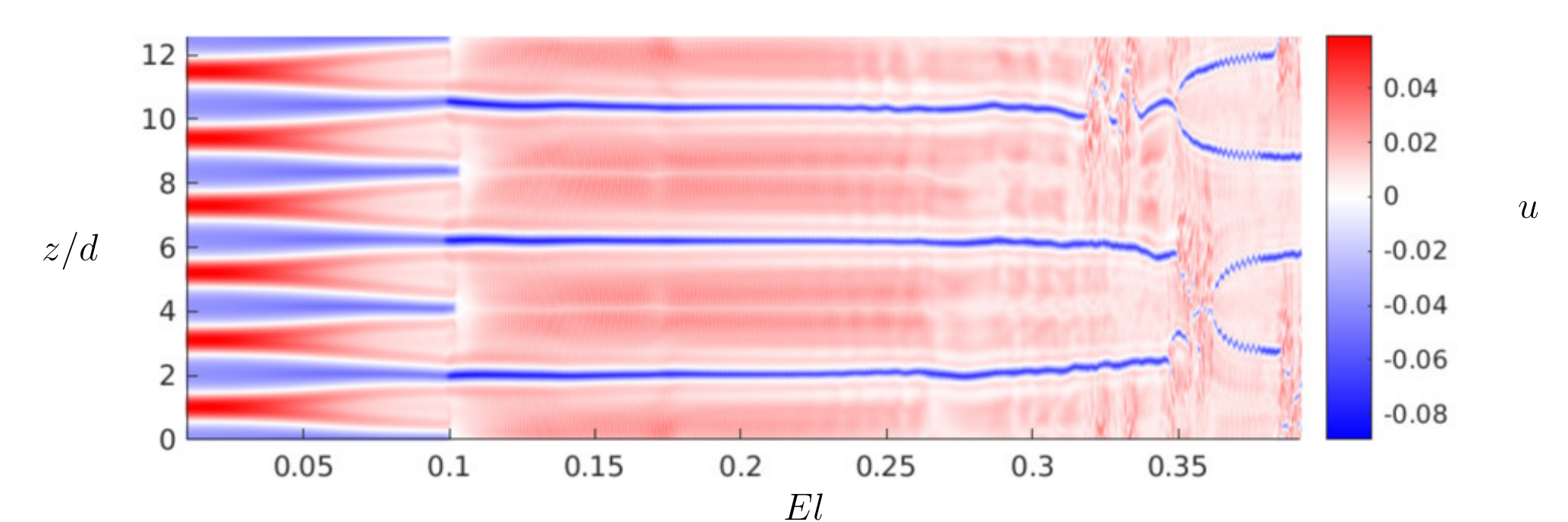} 
  \caption{(Color online) Space-time diagram showing the magnitude of $u$ at the mid-gap along the $z$ direction in a simulation where the $El$ number was slowly decreased with time at the same rate as it was increased in the simulations previously presented ($El = 10^{-3}t/Re$). The parameters used in this simulation are the same as those used in the simulation shown in the section~\ref{sec:to_VMST}. The initial condition corresponds to the state obtained in the simulation where $El$ was slowly increased with time for $El = 0.392$ (indicated by an orange dashed line in the panel (a) of the figure~\ref{fig:spacetime_El}). The flow states observed when $El$ decreases differ from those obtained when $El$ increases, evidencing the existence of hysteresis. Red and blue areas indicate outflows and inflows respectively. Note that periodic boundary conditions are used in $z$.}
  \label{fig:hysteresis}
\end{figure}

\noindent The strong dependence of the onset of the VMS events on the initial condition implies that these are highly nonlinear phenomena. It 
is thus rational to expect that hysteretic behaviour will be observed in the simulations if the control parameter 
is varied in the reversed direction, i.e. if the $El$ number is decreased with time.  To examine this possibility, 
we have conducted a simulation where $El$ has been decreased with time at the same rate as it was increased in the previous simulations. 
The simulation was initiated from the flow state obtained at $El = 0.392$ in the simulation presented in the section \ref{sec:to_VMST} 
(indicated by a dashed line in the panel (a) of the figure~\ref{fig:spacetime_El}) and the same parameter values as in the section 
\ref{sec:to_VMST} were used. To facilitate the description, we denote the simulation in which $El$ increases (decreases) 
with time as forwards (backwards) simulation. Figure~\ref{fig:hysteresis} shows the space-time diagram corresponding to the backwards simulation. 
It becomes apparent from the comparison between this figure and the space time diagram of the forwards simulation 
(panels (a) and (b) of the figure~\ref{fig:spacetime_El}) that there is a strong hysteresis. The flow in the backwards simulation eventually 
returns to the initial state of the forwards simulation (a Taylor vortex flow with 6 pairs of vortices), but it follows a completely different 
path, characterized by VMS events and flow states that differ from those observed in the forwards simulation. It is worth noting that 
the initial cascade of merging events that precede the onset of chaotic motion in forwards simulations is absent in the backwards simulation. 
This reflects the irreversible character of the symmetry-breaking processes that take place over this initial stage of the VMS regime. 
Another notable difference is observed in the transition between the centrifugally and elastically dominated regimes. This occurs at a lower 
$El$ value ($El \approx 0.09$) than in the forwards simulations ($El \approx 0.125$) and not only entails a sudden change in the strength of the vortices,  
but also a change in their number (from 3 vortex pairs in the elastically dominated regime to 6 vortex pairs in the centrifugally dominated regime). 
Once the centrifugally dominated regime is achieved, flow states obtained in the forwards and backwards simulations for the same $El$ values become 
identical. This observation reflects the linear nature of the centrifugal instability and evidence that the hysteretic effects observed 
in the simulations are solely due to the subcritical character of the DW.\\  

\begin{figure}
  \begin{center}
    \includegraphics[scale=0.4]{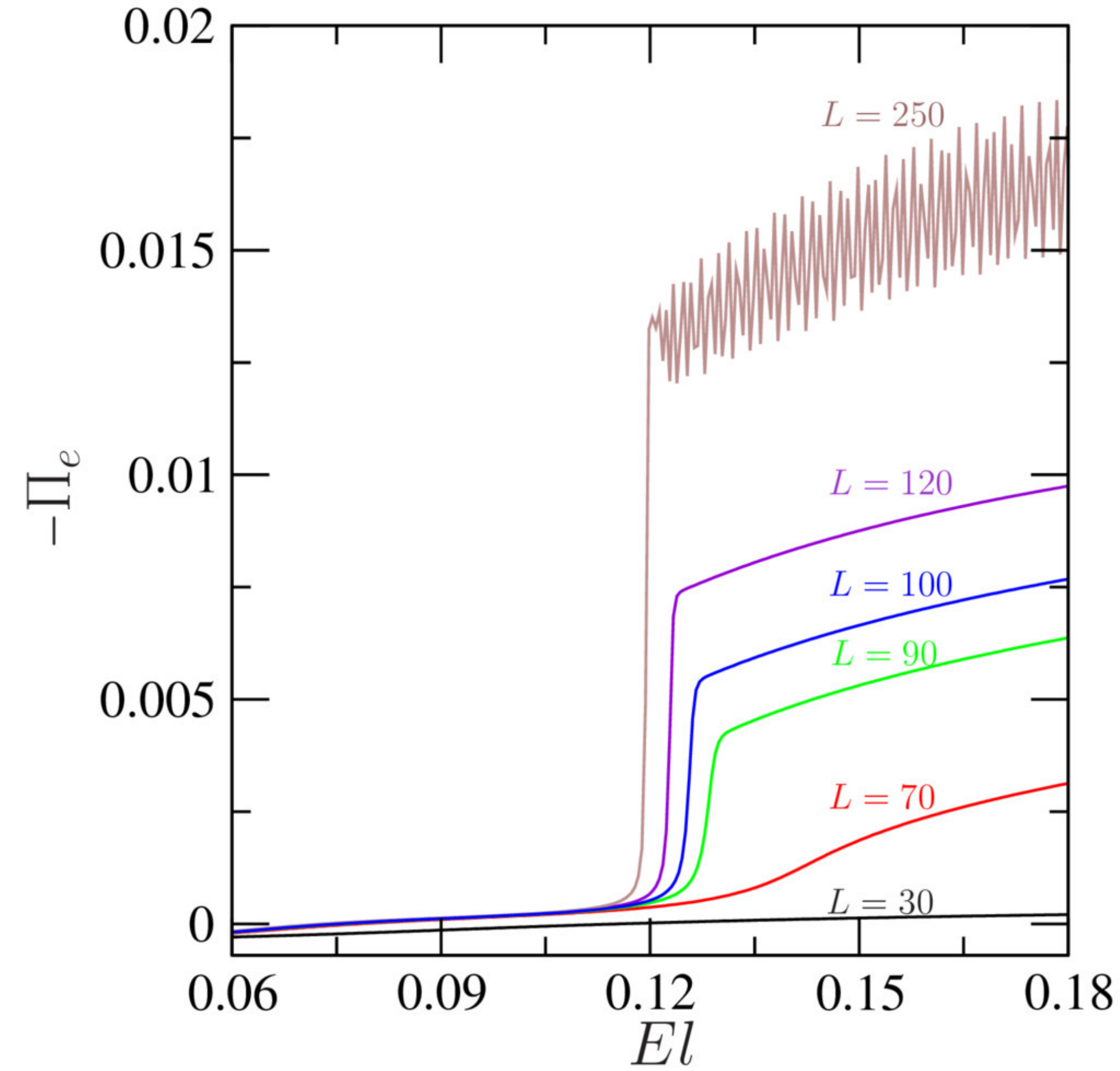}  
  \end{center}
  \caption{ (Color online) Contribution of the elastic stresses to the integral kinetic energy budget $\Pi_e$ as a function of the 
  maximum polymer extension $L$ in the range of $El$ values where the transition between the centrifugally and elastically dominated regimes takes 
  place.}
  \label{fig:elastic_inst_L}
\end{figure}

\noindent We next examine whether the occurrence of VMS events depends on the extensibility of the polymers used in the simulations. To this end, 
we have conducted a set of simulations where the maximum polymer extension $L$ was varied between $30$ and $250$ while keeping the other parameters as in 
section~\ref{sec:to_VMST}. It has been found that the choice of $L$ has an important impact on the characteristics of the elastic instability. This 
is illustrated in the figure~\ref{fig:elastic_inst_L}, which shows the contribution of the polymers to the integral energy budget ($\Pi_e$) in the range of $El$ values 
where the elastic instability takes place. As shown earlier in the figure~\ref{fig:eneg_balance}, the onset of the elastic instability leads 
to a marked increase in the value of $\Pi_e$, which replaces the energy production due to inertia ($\mathcal{P}$) as the leading term that balances the 
viscous dissipation ($\epsilon$). This characteristic is absent in simulations where the extensibility of the polymers is low (see $L = 30$ case in the figure). 
In these simulations, $\Pi_e$ increases very gradually with increasing $El$ and remains negligible with respect to $\mathcal{P}$ and $\epsilon$ for the entire range of $El$ numbers investigated (up to $El = 0.5$). This implies that at these elasticity levels the elastic instability does not occur in these cases. 
As a result, the VMS dynamics does not take place and the flow at high $El$ values is characterized by elastically modified Taylor vortices (not shown). 
A clear increase in $\Pi_e$ consistent with the emergence of an elastic instability is observed in the simulations when these are performed 
using $L \gtrapprox 70$. The $El$ threshold at which the instability sets in decreases slightly with increasing $L$ (from $El \approx 0.135$ for $L=70$ to 
$El \approx 0.115$ for $L=250$). The transition between the centrifugally and elasticity dominated regimes is initially rather smooth (see L = 70 case) 
but it becomes increasingly sharp as $L$ increases. As the transition gets sharper the magnitude of $\Pi_e$ increases substantially, leading to increasingly 
strong vortices  and causing the emergence of spatio-temporal dynamics right after the transition in cases where very large extensibility is considered 
(see L=250 case, where oscillations in the value of $\Pi_e$ arise simultaneously with the transition).\\

\begin{figure}
        \includegraphics[width=\linewidth]{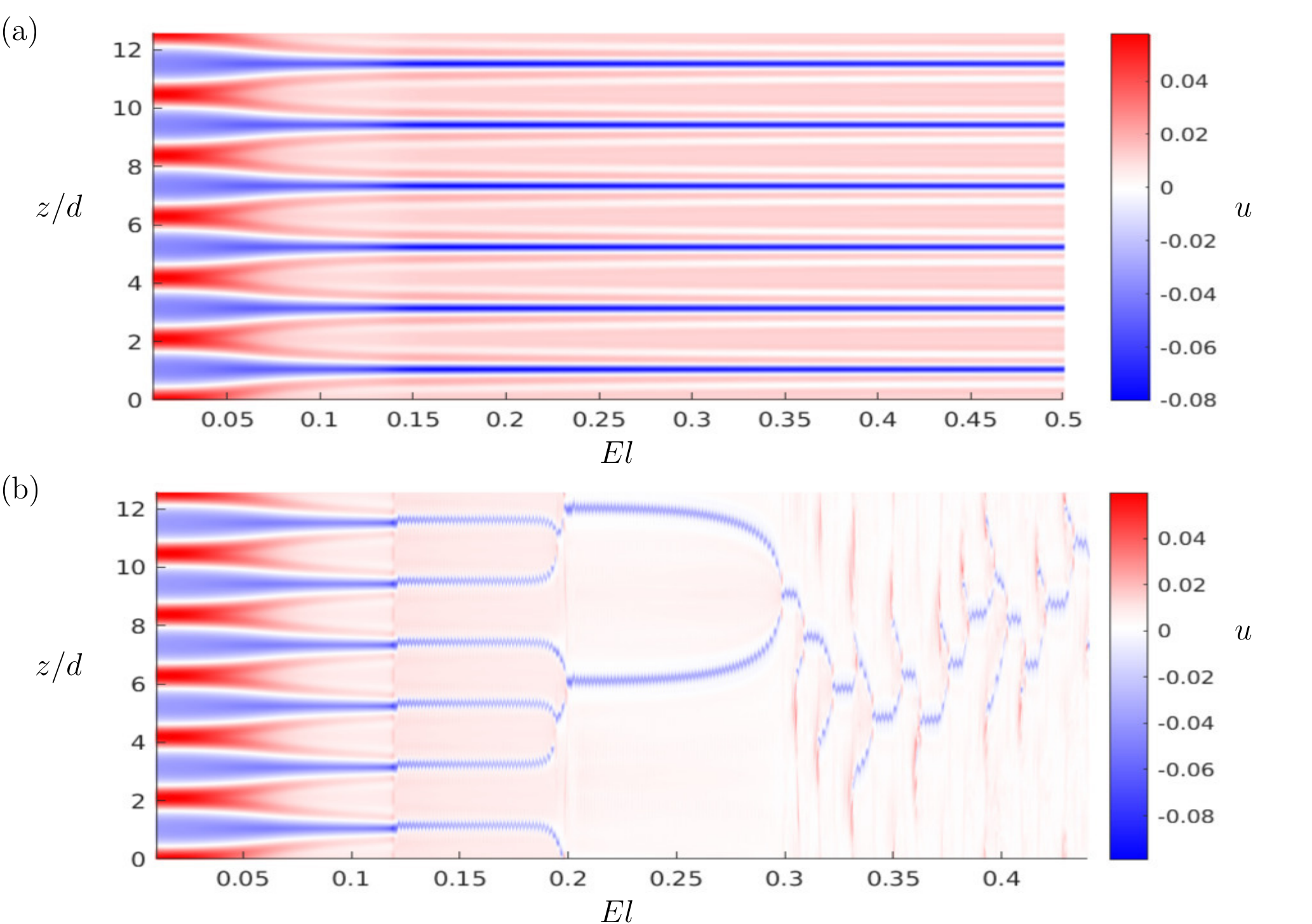} 
    \caption{ (Color online) Space-time diagram showing the magnitude of $u$ at the mid-gap along the $z$ direction in simulations where the maximum polymer extension 
  was set to $L=70$ (a) and $L=250$ (b). The rest of parameters were kept as in section~\ref{sec:to_VMST}. Red and blue areas indicate outflows and inflows 
  respectively.}
  \label{fig:L_effect_space_time}
\end{figure}

\noindent The onset of the VMS dynamics (which has been observed in the simulations where $L \geq 90$) also takes place at smaller values of $El$ as $L$ increases. 
It is interesting that although the elastic instability is observed for $L = 70$, the space time diagram of this simulation (shown in the panel (a) of the 
figure~\ref{fig:L_effect_space_time}) does not show any sign of spatio-temporal complexity. This might reflect that the emergence of VMS events requires 
elasticity levels higher than those simulated here when $L$ is close to the critical value for which the elastic instability emerges. The most striking 
difference in the characteristics of the VMS regime with respect to those observed in the previous simulations occurs when highly extensible polymers are used. 
Here, after the initial cascade of merging events is accomplished, the dynamics is characterized by a sequence of VMS events that exhibits a clear periodicity 
(see the panel (b) of the figure~\ref{fig:L_effect_space_time}, which shows the space time diagram for the simulation using $L = 250$). These structures are 
closely reminiscent of the flame like patterns observed in previous studies~\citep{ThoSurKho06,thomas_khomami_sureshkumar_2009,liu_khomami_2013}, 
with the difference that in these studies the flow was non-axisymmetric and the flame-like dynamics was superposed with a rotating wave, whereas in the present 
study the flow is axisymmetric and therefore the rotating wave is absent.\\ 

\begin{figure}
      \includegraphics[width=\linewidth]{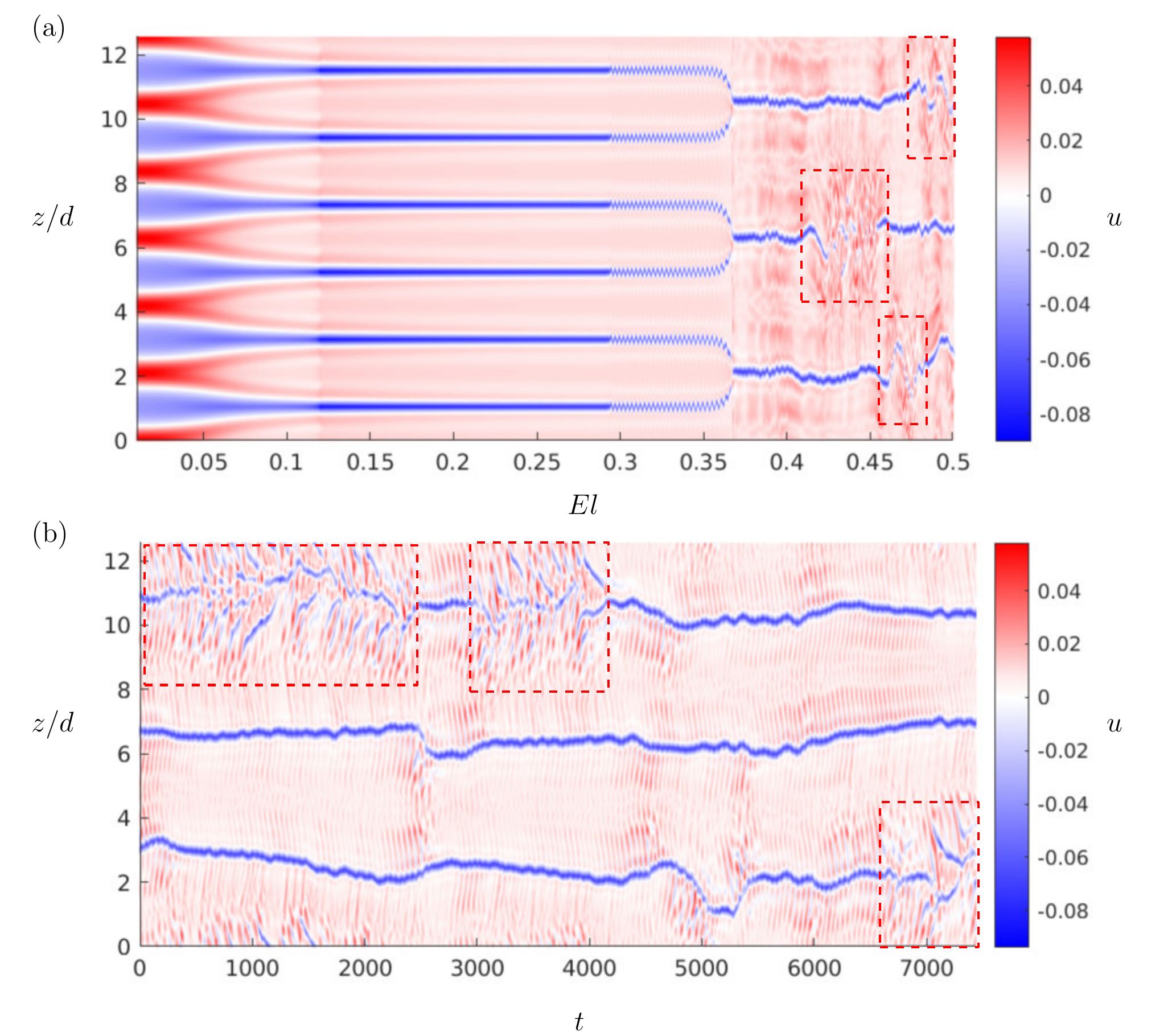} 
  \caption{ (Color online) Example of the dynamics observed in the simulations when $\alpha$ (i.e. the rate of increase in $El$) is 
  larger than $1.25 \cdot 10^{-3}$. (a) Space-time diagram showing the magnitude of $u$ at the mid-gap along the $z$ direction in a simulation 
  where $\alpha = 1.6 \cdot 10^{-3}$. The rest of parameters are as in section~\ref{sec:to_VMST}.  (b) Space time plot obtained when $El$ is held constant at $0.5$ in the simulation shown in (a).
  The dashed (red) rectangles demarcate regions of transient chaotic dynamics. Red and blue areas indicate outflows and inflows respectively.}
  \label{fig:alpha_effect}
\end{figure}

\begin{figure}
 
      \includegraphics[width=\linewidth]{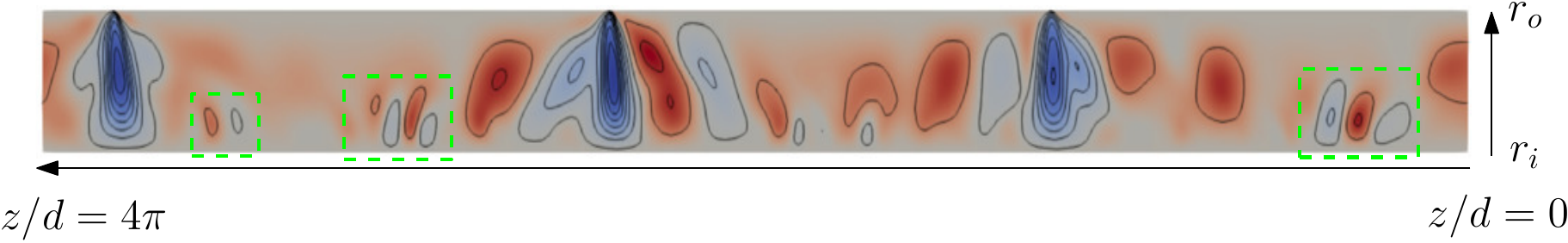} 
   \caption{ (Color online) Color map of $u$ illustrating the instantaneous flow structure at $El = 0.5$ in a simulation where $\alpha = 5\cdot 10^{-3}$. 
  Positive (negative) values of $u$ are shown as red (blue), whereas $u=0$ is shown as gray. There are $10$ contours evenly distributed in $u \in [-0.065,0.015]$. 
  The system is shown rotated by $90$ degrees in the counterclockwise direction. Dashed (green) rectangles are used to highlight the small scale vortices 
  that emerge near the inner cylinder. }
  \label{fig:alpha_effect2}
\end{figure}

\noindent We have finally investigated the effect of varying the rate at which $El$ is increased in the simulations. The increase of the $El$ value with 
time ($El = \alpha t/Re$) can be understood as a continuous perturbation that is imposed on the system, where the parameter $\alpha$ (the rate of increase) 
regulates the amplitude of the perturbation. We have conducted simulations varying $\alpha$ between $7.5 \cdot 10^{-4}$ and $5 \cdot 10^{-3}$, while keeping 
the other parameters as in section~\ref{sec:to_VMST}. The VMS regime (with characteristics similar to those reported in the previous subsections) was found 
in the simulations where $7.5 \cdot 10^{-4}  \lessapprox \alpha \lessapprox  1.25 \cdot 10^{-3}$. However, when $\alpha$ was set to higher 
values, the VMS dynamics did not occur. The panel (a) of the figure~\ref{fig:alpha_effect} illustrates the space time plot as a function of $El$ for a 
simulation where $\alpha = 1.6 \cdot 10^{-3}$. As seen, the variation of the vortex pattern with increasing $El$ is initially analogous to that observed 
when $\alpha = 1 \cdot 10^{-3}$ (panel (a) of the figure~\ref{fig:spacetime_El}). The transition between the centrifugally and elasticity dominated regimes 
taking place at $El \approx 0.12$ is clearly identified by the sudden change of the color intensity of the vortices. Also as in the simulation for 
$\alpha = 1 \cdot 10^{-3}$, the vortex pattern becomes unstable at $El \approx 0.29$, resulting in small axial oscillations of the DW. These oscillations 
persist until several merging events take place simultaneously and break the axial symmetry of the flow pattern (which occurs at $El \approx 0.36)$. 
However, the flow regime that emerges after the symmetry breaking process differs from the VMS regime. The DW remain at approximately the same 
axial positions and their number does not change with increasing $El$, nor with time if $El$ is held constant, as shown in the panel (b) of the same 
figure. Similar to the VMS regime, localized transient chaotic dynamics is often observed in this flow regime (see regions enclosed by the dashed (red) rectangles). 
As shown in the figure~\ref{fig:alpha_effect2}, a particular feature of the flow structure in this flow regime is the emergence of small scale vortices near the inner cylinder 
(see regions demarcated by the dashed (green) rectangles). These vortices are consistent with the elastic G\"ortler vortices identified by~\cite{song_teng_liu_ding_lu_khomami_2019} at higher 
Reynolds numbers. The finding of a flow regime distinct from the VMS regime in these simulations reflects a well known feature of viscoelastic TCF: 
the coexistence of various flow regimes for the same values of the control parameters. The simulations may converge to one regime or the other depending on the amplitude 
and type of the perturbations imposed. Our study shows that to capture the VMS in the simulations, the amplitude of the perturbation cannot be too large, 
otherwise it is replaced by the flow regime just described.

\subsection{Variation of the angular momentum transport with increasing the fluid's elasticity level}\label{sec:momentum}

\noindent An important question raised by the observations above is why the dynamics of the distinct vortex pairs decouple when the polymers elasticity exceeds a certain threshold. It is well known that in two-dimensional vortex systems the conservation of angular momentum imposes strong restrictions on the motion of the vortices and the mean separation among them remains generally nearly constant~\citep{batchelor_2000,Aref83}. It may therefore appear surprising that vortex pairs in the VMS regime can freely move through the system, either toward each other or away from each other, without drastic changes in the system's energy. To explain this seeming inconsistency, it is instructive to examine the impact of increasing $El$ in the different contributions to the flux of angular momentum ($J^{\omega}$) across the annular gap.  In a viscoelastic fluid flow, $J^{\omega}$ can be split in three terms,
\begin{equation}\label{eq:Jw_contrib}
  J^{\omega} = r^3[\underbrace{\overline{u\omega}}_{J^{\omega}_c} - \underbrace{\frac{\beta}{Re} \partial_r\overline{\omega}}_{J^{\omega}_d} - \underbrace{\frac{(1-\beta)}{Re}\frac{\overline{T_{r\theta}}}{r}}_{J^{\omega}_p}],
\end{equation}
where $\omega = v/r$ is the angular velocity and the bar symbol indicates averaging over the axial direction (for $El$ values where the flow is steady) or over the axial direction and time (when the flow is non-steady or chaotic). Here, $J^{\omega}_c$ denotes the convective transport of angular momentum, which is associated with the vortices, $J^{\omega}_d$ is the diffusive transport due to viscosity and $J^{\omega}_p$ is the angular momentum transport caused by polymer stresses. \\

\begin{figure}\setlength{\piclen}{\linewidth}
  \begin{center}
      \includegraphics[width=\piclen]{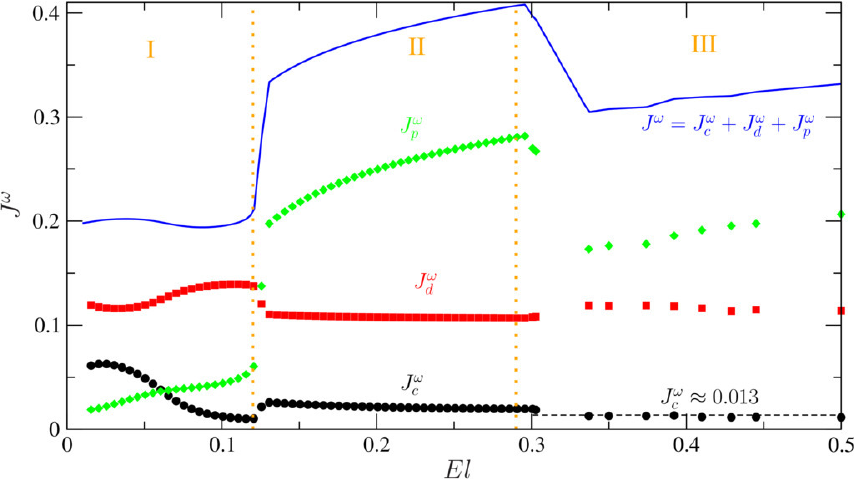}    
  \end{center}
  \caption{ (Color online) Variation of the axially and time averaged angular momentum current ($J^{\omega}$) and its contributing terms, $J^{\omega}_c$, $J^{\omega}_d$ and $J^{\omega}_p$, obtained near mid-gap, as $El$ increases. The (orange) vertical dotted lines demarcate the different flow regimes observed in the present study: 
  I. the centrifugally dominated regime, II. the elastically dominated regime characterized by steady patterns of equally spaced diwhirls, III. the elastically 
  dominated regime characterized by spatio-temporal dynamics.}
  \label{fig:momentum}
\end{figure}

\noindent Although the above equation was originally derived for turbulent flow~\citep{eckhardt_grossmann_lohse_2007,song_teng_liu_ding_lu_khomami_2019}, it is straightforward to show that it also applies to steady vortex flow (see appendix~\ref{appB} for a step by step derivation under steady and axisymmetric conditions). Hence, it can be used to study the variation in the contributions of the different transport mechanisms as $El$ increases from the Newtonian limit up to the the largest value simulated within the VMS regime. This is shown in the figure~\ref{fig:momentum} for a radial location near the mid-gap using the data obtained from the simulation presented in the section~\ref{sec:to_VMST} (similar analyses for other simulations are shown in the appendix~\ref{sec:appC}). Since the dynamics in the VMS regime is chaotic and $J^{\omega}$ is a fluctuating quantity, several additional simulations at constant $El$ had to be performed in order to obtain meaningful values of $J^{\omega}$ and its contributing terms in this flow regime. The initial conditions for these simulations were flow states obtained in the simulation with slowly varying $El$. Starting from these solutions, the $El$ values were fixed and the chaotic flow was simulated in each case for $20000$ advective time units. The momentum fluxes were then calculated by averaging over this long time period.\\ 

\noindent As seen in the figure, three clear stages can be distinguished in the behaviour of the angular momentum fluxes as $El$ increases, which are consistent with the different flow regimes identified throughout our study. In the first stage, coinciding with the centrifugally dominated regime, $J^{\omega}$ remains nearly constant. While its main contribution stems from the diffusive transport ($J^{\omega}_d$), the convective flux ($J^{\omega}_c$) also plays an important role close to the Newtonian limit ($El \to 0$). However, due to the dissipative nature of the polymers activity in this flow regime, the contribution of the vortices to $J^{\omega}$ decays with increasing $El$ and it is gradually replaced by the angular momentum flux due to the polymer stresses ($J^{\omega}_p$). Note that although it may seem surprising that the contribution of $J^{\omega}_d$ exceeds that of $J^{\omega}_c$ in a Newtonian (or low $El$) vortex flow, this happens because the simulation is conducted at a Reynolds number ($Re=95$) which is very close to the onset of the Taylor vortices ($Re = 89$). Here, the vortices are still weak and the momentum transport is dominated by molecular diffusion (as in the laminar regime). As $Re$ increases, the contribution of $J^{\omega}_c$ near the mid-gap becomes increasingly large and the contribution of $J^{\omega}_d$ decreases, so that the former eventually dominates the momentum transport in this regime (not shown). The onset of the elasticity dominated regime ($El \sim 0.12$) is accompanied by an abrupt increase in $J^{\omega}_p$, which becomes the leading contribution to the angular momentum flux. Moreover, the magnitude of $J^{\omega}_p$ keeps increasing as  $El$ increases, leading to a substantial increase in $J^{\omega}$ with respect to that in the centrifugally dominated regime. The convective and diffusive fluxes, on the other hand, exhibit a slight increase and decrease, respectively, when the elastic instability sets in, and subsequently decrease very gradually with increasing $El$. The onset of the spatio-temporal dynamics ($El \sim 0.29$) brings a significant drop in the total angular momentum flux. This is mainly associated with an initial decay in $J^{\omega}_p$ that takes place during the initial cascade of merging events where the different vortex pairs become fully decoupled. Despite its initial decrease, $J^{\omega}_p$ is still the main contribution to $J^{\omega}$, and after the initial phase of merging events, its magnitude increases with increasing $El$ over the entire VMS regime. The convective contribution $J^{\omega}_c$ also decays during the initial phase of merging events. However, it appears to fluctuate around a constant value, $J^{\omega}_c \approx 0.013$, with increasing $El$. A similar behaviour is observed in $J^{\omega}_d$, which increases initially with the emergence of the VMS dynamics but remains subsequently nearly constant as $El$ increases.\\

\noindent The analysis of the angular momentum fluxes just presented allows us to propose an answer to the question posed at the beginning of this section. As shown, polymer stresses are very efficient in transporting angular momentum (provided that the level of elasticity in the working fluid is sufficiently large) and so the contribution of the vortices in this regard, which is essential in the Newtonian case, is only marginal in the viscoelastic case. The amount of momentum carried by the vortices becomes increasingly small as $El$ increases until it eventually reaches a nearly constant level, $J^{\omega}_c \approx 0.013$, in all simulations where the VMS is found. On the basis of this observation, we suggest that when the angular momentum carried by the vortices drops to this level, the constraints imposed by the angular momentum conservation on the vortices break and these may decouple from the rest of the system. This limit would mark the beginning of the VMS dynamics and could also be interpreted as the minimum amount of angular momentum that the DW must carry to form a pattern of steady vortices. The latter interpretation offers an explanation to the question of why the DW do not merge at lower $El$ values. As noted in section~\ref{sec:to_VMST}, arrangements of equally spaced, steady DW have not been so far experimentally reported. In fact, it has been inferred from the experiments that two DW coalesce when the distance between them is lower than $5d$, a characteristic that would render the formation of these arrangements of DW unfeasible. The reason for this apparent contradiction may lie in the fact that these experiments were conducted at low values of $Re$, where the flow in the Newtonian case is centrifugally stable (i.e. molecular diffusion suffices to transport angular momentum). The amount of angular momentum carried by the DW at these low $Re$ is expected to be very small and hence it is reasonable to assume that it might be below the threshold required to observe these arrangements. Another important remark concerning the angular momentum fluxes in the VMS regime is that the two Newtonian contributions ($J^{\omega}_c$ and $J^{\omega}_d$) remain nearly constant over the entire regime. The increase in the angular momentum flux taking place in this flow regime is thus only due to the contribution of the polymer stresses, which continues increasing with increasing $El$. This circumstance strongly suggests that the dynamics in the VMS regime is fully dominated by  elastic effects (as already noted in the experimental study by \cite{LaCaGiBa20}) and raises the question about a possible relationship between flow states in the VMS regime and those driven by pure elastic instabilities in the inertialess limit.\\

\section{Conclusions}\label{sec:conclusion}

\noindent We have presented numerical simulations of viscoelastic TCF aimed at elucidating recent experimental observations of an elastically-induced chaotic dynamics governed by a successive merging and splitting of vortices,  known as the vortex merging and splitting (VMS) transition to elasto-inertial turbulence \citep{LaCaGiBa20}. Unlike the experiments, where this regime was achieved by increasing $Re$ while keeping a constant $El$ level, in the present study we have fixed $Re$ to $95$ (a value that falls within the centrifugally unstable regime of TCF) and have increased $El$ progressively starting from a Taylor vortex flow pattern in the Newtonian limit ($El = 0$). This different protocol have allowed us to investigate the transformation and instabilities that the axisymmetric vortex flow pattern undergoes as $El$ increases and the VMS regime is achieved.\\  

\noindent Our simulations show that, unlike other transition scenarios to elasto-inertial turbulent states (e.g. the transition via flame patterns~\citep{latrache2021transition} or the transition driven by defects~\citep{LaAbCruMu16}), the transition to the VMS regime does not involve any thee-dimensional instability and it is associated with instabilities of an axisymmetric vortex pattern which are induced solely by elastic effects. The centrifugal instability mechanism giving rise to the well known Taylor vortices is found to persist at low-to-moderate values of $El$. Nevertheless, polymers induce strong dissipative effects in this flow regime, causing drastic quantitative and structural changes in the vortex pattern. These changes are particularly pronounced at low $El$ values, thereby evidencing the immediate and dramatic impact that the addition of polymers has on the flow characteristics.\\

\noindent A key factor to explain the complex spatio temporal dynamics observed at high $El$ values is the transformation the flow undergoes at $El \approx 0.12$. Above this $El$ threshold, polymers inject energy into the flow through the elastic stresses and the centrifugal mechanism inducing the vortex pattern is replaced by an elastic mechanism. The result of the elastic instability is a steady vortex pattern where the vortex pairs are identified as diwhirls: a type of vortical structure similar to a dipole which is characterized by a strong asymmetry between inflows and outflows. While these structures have been well documented in the literature~\citep{GroiStei97,LaBru01,ThoSurKho06,thomas_khomami_sureshkumar_2009}, there are some important differences between the state found in our simulations and those previously reported. The first important distinction is the pathway we have followed to find these structures. Previous experiments and simulations suggested that diwhirls could only emerge as the inner cylinder speed is decreased (i.e. as $Re$ decreases). In our simulations, however, diwhirls appear at a constant value of $Re$ as $El$ increases, reflecting that deceleration is not a necessary condition to observe these structures. A second and more important distinction is the spatial arrangement of the diwhirls. Previous studies on diwhirls were conducted at low $Re$ values (in centrifugally stable regimes) where it was found that nearby diwhirls always approach each other and coalesce into a single entity. As a result, after a certain time diwhirls usually appear as solitons in these flow regimes. Our simulations reveal that, in a centrifugally unstable regime, diwhirls do not necessarily merge when they are close to each other, and may appear for a wide range of  $El$ values as a steady vortex pattern.\\ 

\noindent Vortex merging and splitting events take place when the dynamics of the distinct diwhirls decouple and these begin to travel freely in the axial direction.  We propose that this dynamical decoupling is possible because as $El$ increases the amount of angular momentum carried by the vortices reaches a marginal level (the angular momentum transport across the gap is mainly due to the polymer stresses). This circumstance permits the vortices break the constraints that conservation of momentum imposes on their motion, thus making it possible for them to be dynamically disconnected. During the onset phase of the VMS regime only merging events are observed, but with further increase in $El$, a complex spatio-temporal dynamics characterized by a series of merging and splitting events, which closely resembles experimental observations, is also found in the simulations. Merging events occur when two diwhirls move in opposite senses towards each other. As they get close to one another, they experience a strong attractive interaction that increases their travelling speeds and accelerates the merging process. Conversely, splitting events occur when newly created vortex pairs move away from each other and get separated by a distance that is sufficiently large so that their mutual interaction is weak. The occurrence of splitting events is always preceded by local regions of transient chaotic motion which result from merging events as a consequence of the redistribution of the kinetic energy released by the diwhirls that are eliminated in these events. A key feature of the VMS regime is the existence of a power law decay range in the power spectrum with a $-3$ exponent. Such decay rate complies with the universal power law spectral decay expected for elasto-inertial turbulence~\citep{Foux03,Ya21} and thus suggests categorizing the VMS regime as a class of the elasto-inertial turbulent states.\\

\noindent Due to the highly nonlinear nature of the diwhirls, changes in the aspect ratio and/or the number of vortex pairs of the initial condition may notable alter the $El$ threshold at which the VMS regime sets in. Specifically, the onset of this regime has been found to vary between $El \approx 0.15$ and $El \approx 0.35$ depending on the aspect ratio and the initial condition used in the simulations. This range of $El$ values is consistent with the elasticity level at which these dynamics have been reported in the experiments, $El \approx 0.22$~\citep{LaCaGiBa20}. The characteristics of the VMS events, as well as the transition towards the VMS regime as $El$ increases, are largely similar in all simulations. The main exception is the regime characterized by the small axial oscillations of the diwhirls, i.e. the standing wave described in the section~\ref{sec:complex}. This regime precedes the VMS dynamics in simulations where the latter occurs at high $El$ values, but it is absent in those where VMS events already occur at moderate values of $El$. In these latter cases, the oscillations are also observed at high $El$ values, but they coexist with the VMS dynamics. Hence, it may be concluded that the emergence of the standing wave is not a necessary condition for the onset of the VMS regime.\\ 

\noindent We have also studied how changes in the maximum polymer extension $L$ and the parameter $\alpha$ (the rate of increase in $El$) modify the outcomes of the simulations. It has been found that if $L$ is too small ($L < 70$), the elastic instability does not set in and consequently the VMS dynamics does not take place. Conversely, if $L$ is too large ($L > 200$), the simulations converge to a different flow regime, which is consistent with the flames-like structures previously reported in the literature~\citep{ThoSurKho06,thomas_khomami_sureshkumar_2009,liu_khomami_2013}. The VMS regime has been found in simulations where the value of $L$ is between $90$ and $200$.  An appropriate choice of $\alpha$ is also essential to capture the VMS in the simulations. This regime has been found in simulations where $7.5 \cdot 10^{-4} \lessapprox \alpha \lessapprox 1.25 \cdot 10^{-3}$. For $\alpha > 1.25 \cdot 10^{-3}$, however, a different flow regime emerges at high $El$ values. VMS events do not occur in this regime (i.e. the number of vortex pairs remain constant) and small scale vortices, consistent with elastic G\"ortler vortices~\citep{song_teng_liu_ding_lu_khomami_2019}, appear near the inner cylinder.\\

\noindent In closing, we would like to note that there are many important open questions about this topic which have not been addressed yet. For example, a detailed study of the VMS transition as $Re$ increases, examining the changes the flow undergoes until it converges to the eventual elasto-inertial state, is missing. A statistical and structural characterization of the elasto-inertial turbulent state has not been done either, which has prevented from any comparisons with other elasto-inertial turbulent states reported in viscoelastic TCF~\citep{liu_khomami_2013,song_teng_liu_ding_lu_khomami_2019,song_lin_liu_lu_khomami_2021} or in other fluid flow systems~\citep{Samanta13,Dubief13}. It is also unclear whether there exist a connection between this elasto-inertial turbulent regime (which requires of pure elastic instabilities to exist) and the pure elastic turbulent regime taking place at vanishing inertia. Finding answers to these and other related questions guarantees that viscoelastic TCF will be an active focus of research in the upcoming years.\\ 

\noindent This work has been supported by the grant PID2020-114043GB-I00 of the Spanish Ministry of Science and Innovation. The author thankfully acknowledges the computer resources at Pirineus and the technical support provided by the Consorci de Serveis Universitari de Catalunya (RES-IM-2022-1-0005).\\

\noindent Declaration of Interests. The authors report no conflict of interest.\\

\appendix

\section{Energy balance equation for steady axisymmetric vortex flow}\label{appA}

\noindent In this section it is shown that an equation for the energy associated with steady and axisymmetric vortices can be derived following a line of reasoning similar to that typically used in the derivation of the turbulent kinetic energy equation. Under conditions of steadiness and axisymmetry (i.e. $\partial_t = 0$ and $\partial_{\theta} = 0$), and using the notation and  non-dimensionalization described in the section~\ref{sec:Problem},  the momentum equations for a viscoelastic TCF read 
\begin{align}
      0 = -u\partial_r u - w\partial_{z}u +\frac{v^2}{r} - \partial_r p + \frac{\beta}{Re}(\nabla^2 u - \frac{u}{r^2}) + \frac{1-\beta}{Re}(\partial_r T_{rr} + \frac{(T_{rr}-T_{\theta\theta})}{r} + \partial_z T_{rz})\label{eq:azim_steady_NV_visco_radial}\\ 
      0 = -u\partial_r v - w\partial_{z}v -\frac{uv}{r} + \frac{\beta}{Re}(\nabla^2 v- \frac{v}{r^2}) + \frac{1-\beta}{Re}(\partial_r T_{r\theta} + \frac{2T_{r\theta}}{r} + \partial_z T_{\theta z})\label{eq:azim_steady_NV_visco_azi}\\
      0 = -u\partial_r w - w\partial_{z}w -\partial_z p + \frac{\beta}{Re}\nabla^2 w + \frac{1-\beta}{Re}(\partial_r T_{rz} + \frac{T_{rz}}{r} + \partial_z T_{zz})\label{eq:azim_steady_NV_visco_axial},
\end{align}
where the Laplacian term is given by
\begin{equation}\label{eq:laplacian}
       \nabla^2 f = \frac{1}{r}\partial_r(r\partial_r f)+\partial_z^2 f
\end{equation}

\noindent We first obtain the  axially averaged momentum equations. To that extent, the velocity, pressure and polymer stress tensor of the vortex flow are decomposed as 
\begin{equation}\label{eq:mean_decomposition}
    \mathbf{v} = \begin{bmatrix}       0 \\ \overline{v} \\ 0 \end{bmatrix} + \begin{bmatrix}       u' \\ v' \\ w' \end{bmatrix}, p = \overline{p} + p', \mathbf{T} = \begin{bmatrix} \overline{T_{rr}}\\ \overline{T_{r\theta}} \\ \overline{T_{rz}} \\ \overline{T_{\theta\theta}} \\ \overline{T_{\theta z}} \\  \overline{T_{zz}} \end{bmatrix}  + \begin{bmatrix} T'_{rr}\\ T'_{r\theta} \\ T'_{rz} \\ T'_{\theta\theta} \\ T'_{\theta z} \\  T'_{zz} \end{bmatrix}
\end{equation}
where the bar symbol is used to indicate that the variables are axially averaged and the prime symbol denotes deviation from the axially averaged value. Note that for a vortex flow pattern these operators satisfy the Reynolds averaged rules (i.e. $\overline{f'} = 0$, $\overline{\overline{f}}= \overline{f}$ and $\overline{\overline{f} f'}=0$). Substituting this decomposition into the momentum equations and averaging over the axial direction, one obtains  

\begin{align}
  0 = -\overline{u'\partial_r u'} - \overline{w'\partial_{z}u'} +\frac{\overline{v}^2}{r} +\overline{\frac{v'^2}{r}} - \partial_r \overline{p} +\frac{1-\beta}{Re}(\partial_r \overline{T_{rr}} +  \frac{\overline{T_{rr}}-\overline{T_{\theta\theta}}}{r}) \label{eq:azim_steady_NV_visco_rad_avg}\\  
      0 = - \overline{u'\partial_r v'} - \overline{w'\partial_{z}v'} -\frac{\overline{u'v'}}{r} + \frac{\beta}{Re}(\frac{1}{r}\partial_r(r\partial_r \overline{v}) - \frac{\overline{v}}{r^2}) + \frac{1-\beta}{Re}(\partial_r \overline{T_{r\theta}} + \frac{2\overline{T_{r\theta}}}{r}) \label{eq:azim_steady_NV_visco_azi_avg}\\
      0 = -\overline{u'\partial_r w'} - \overline{w'\partial_{z}w'}  + \frac{1-\beta}{Re}(\partial_r \overline{T_{rz}} + \frac{\overline{T_{rz}}}{r}) \label{eq:azim_steady_NV_visco_axial_avg}
\end{align}      
Using the product rule for derivatives and the incompressibility condition, the above equations can be rewritten as  
\begin{align}
         0 = -\partial_r \overline{u'u'} - \frac{\overline{u'u'}}{r}  +\frac{\overline{v}^2}{r} +\frac{\overline{v'v'}}{r} - \partial_r \overline{p} +\frac{1-\beta}{Re}(\partial_r \overline{T_{rr}} +  \frac{\overline{T_{rr}}-\overline{T_{\theta\theta}}}{r}) \label{eq:azim_steady_NV_visco_rad_avg_final}\\  
      0 = -\partial_r \overline{u'v'} - \frac{2\overline{u'v'}}{r} + \frac{\beta}{Re}(\frac{1}{r}\partial_r(r\partial_r \overline{v}) - \frac{\overline{v}}{r^2}) + \frac{1-\beta}{Re}(\partial_r \overline{T_{r\theta}} + \frac{2\overline{T_{r\theta}}}{r})\label{eq:azim_steady_NV_visco_azi_avg_final}\\
      0 = -\partial_r \overline{u'w'} - \frac{\overline{u'w'}}{r}  + \frac{1-\beta}{Re}(\partial_r \overline{T_{rz}} + \frac{\overline{T_{rz}}}{r})\label{eq:azim_steady_NV_visco_axial_avg_final}
\end{align}      
which is the final form of the axially averaged momentum equations.\\ 

\noindent We now multiply~\cref{eq:azim_steady_NV_visco_radial,eq:azim_steady_NV_visco_azi,eq:azim_steady_NV_visco_axial} by the velocity field and average over the axial direction 
to obtain an equation for the total kinetic energy. The resulting equation is
\begin{multline}\label{eq:energy_total}
        0 = - \overline{u \partial_r u u} - \overline{w \partial_z u u} - \overline{u \partial_r v v} - \overline{w \partial_z v v}
    - \overline{u \partial_r w w} - \overline{w \partial_z w w} - \overline{\partial_r p u} - \overline{\partial_z p w}\\
    + \frac{\beta}{Re}( \overline{u \nabla^2 u} - \overline{\frac{u^2}{r^2}} + \overline{v \nabla^2 v} - \overline{\frac{v^2}{r^2}} + \overline{w \nabla^2 w})  + \frac{1-\beta}{Re}(\overline{u \partial_r T_{rr}} + \overline{u \frac{(T_{rr}-T_{\theta\theta})}{r}} + \overline {u \partial_z T_{rz}} \\
    + \overline{v \partial_r T_{r\theta}} + \overline{v \frac{2T_{r\theta}}{r}} + \overline{v \partial_z T_{\theta z}} + \overline{w \partial_r T_{rz}} + \overline{w \frac{T_{rz}}{r}} + \overline{w \partial_z T_{zz}})
\end{multline}
Introducing the decomposition in \cref{eq:mean_decomposition} and subtracting \cref{eq:azim_steady_NV_visco_azi_avg_final} multiplied by $\overline{v}$ (i.e. the equation for the axially averaged kinetic energy), an equation for the kinetic energy of the vortices is obtained
\begin{multline}\label{eq:energy_vortices}
        0 = - \overline{u' \partial_r u' u'} - \overline{w' \partial_z u' u'} - \overline{u' \partial_r \overline{v} v'} - \overline{u' \partial_r v' \overline{v}} 
    -\overline{u' \partial_r v' v'}    - \overline{w' \partial_z v' \overline{v}} - \overline{w' \partial_z v' v'} 
    - \overline{u' \partial_r w' w'} \\
    - \overline{w' \partial_z w' w'} - \overline{\partial_r p' u'} - \overline{\partial_z p' w'}
    +\overline{v} \partial_r \overline{u'v'} + \frac{2\overline{v}\overline{u'v'}}{r}  
    + \frac{\beta}{Re}( \overline{u' \nabla^2 u'} - \overline{\frac{u'^2}{r^2}} + \overline{v' \nabla^2 v'} - \overline{\frac{v'^2}{r^2}} + \overline{w' \nabla^2 w'})\\
    + \frac{1-\beta}{Re}(\overline{u' \partial_r T'_{rr}} + \overline{u' \frac{(T'_{rr}-T'_{\theta\theta})}{r}} + \overline {u' \partial_z T'_{rz}} + \overline{v' \partial_r T'_{r\theta}} + \overline{\frac{2v'T'_{r\theta}}{r}} + \overline{v' \partial_z T'_{\theta z}} + \overline{w' \partial_r T'_{rz}} + \overline{w' \frac{T'_{rz}}{r}} \\
    + \overline{w' \partial_z T'_{zz}})
\end{multline}
With some manipulation (using again the product rule for derivatives and the incompressibility condition), the above equation can be rewritten in the form
\begin{multline}\label{eq:final_equation_no_integral}
        0 = -\frac{1}{r} \partial_r \bigl[ r\bigl( \frac{1}{2} (\overline{u'u'u'} + \overline{u'v'v'} 
          + \overline{u'w'w'}) + \overline{p' u'} - \frac{\beta}{2Re} \partial_r\left(\overline{u'u'} + \overline{v'v'} + \overline{w'w'}\right) \\
          -\frac{1-\beta}{Re} (\overline{u' \partial_r T'_{rr}} + \overline{v' \partial_r T'_{r\theta}} + \overline{w' \partial_r T'_{rz}})\bigr)\bigr] - \overline{u'v'} (\partial_r \overline{v} - \frac{\overline{v}}{r}) \\       
    - \frac{\beta}{Re}( \overline{\partial_r u' \partial_r u'} + \overline{\partial_z u' \partial_z u'} + \overline{\partial_r v' \partial_r v'} + \overline{\partial_z v' \partial_z v'} + 
    \overline{\partial_r w' \partial_r w'} + \overline{\partial_z w' \partial_z w'} + \overline{\frac{u'^2}{r^2}} + \overline{\frac{v'^2}{r^2}})\\ 
    - \frac{1-\beta}{Re}(\overline{\partial_r u' T'_{rr}} + \overline{\frac{(u'T'_{\theta\theta})}{r}} + \overline {\partial_z u' T'_{rz}} + \overline{\partial_r v'  T'_{r\theta}} + \overline{\frac{v' T'_{r\theta}}{r}} + \overline{\partial_z v' T'_{\theta z}} + \overline{\partial_r w' T'_{rz}}  \\
    + \overline{\partial_z w' T'_{zz}})
\end{multline}
Finally, to obtain the volume average kinetic energy of the vortices that is presented in the figure~\ref{fig:eneg_balance},~\cref{eq:final_equation_no_integral} is integrated over the radial direction. In doing so, the first term of the equation (i.e. the radial derivative of the quantity between brackets), which represents energy transport due to the various transport mechanisms at play, becomes zero and the integral kinetic energy equation reads as 
\begin{multline}\label{eq:final_equation_integral}
        0 = -\int_{0}^{1} (\overline{u'v'} (\partial_r \overline{v} - \frac{\overline{v}}{r})) rdr \\
    -\frac{\beta}{Re}\int_{0}^{1} (\overline{\partial_r u' \partial_r u'} + \overline{\partial_z u' \partial_z u'} + \overline{\partial_r v' \partial_r v'} + \overline{\partial_z v' \partial_z v'} + 
    \overline{\partial_r w' \partial_r w'} + \overline{\partial_z w' \partial_z w'} + \overline{\frac{u'^2}{r^2}} + \overline{\frac{v'^2}{r^2}}) rdr\\ 
    -\frac{1-\beta}{Re}\int_{0}^{1} (\overline{\partial_r u' T'_{rr}} + \overline{\frac{(u'T'_{\theta\theta})}{r}} + \overline {\partial_z u' T'_{rz}} + \overline{\partial_r v'  T'_{r\theta}} + \overline{\frac{v' T'_{r\theta}}{r}} + \overline{\partial_z v' T'_{\theta z}} + \overline{\partial_r w' T'_{rz}}  \\
    + \overline{\partial_z w' T'_{zz}}) rdr
\end{multline}
The equation above is the same as equation~\eqref{eq:eneg_balance} but written in terms of its non-zero components. The first integral corresponds to the production of kinetic energy due to deviations of the velocity field from the axially averaged velocity ($\mathcal{P}$), the second integral is the viscous dissipation of the kinetic energy ($\epsilon$) and the third integral represents the contribution of the polymers ($\Pi_e$), which may be a production or a dissipation term depending on the fluid's elasticity.

\section{The angular velocity current for steady axisymmetric vortex flow}\label{appB}

\noindent The equation~\eqref{eq:Jw_contrib} used to decompose the radial flux of the angular velocity $\omega$ as a function of the contributions of the diffusive, convective and elastic transport mechanisms was originally derived by~\cite{eckhardt_grossmann_lohse_2007} for fully turbulent Newtonian TCF and later extended to the viscoelastic case by~\cite{song_teng_liu_ding_lu_khomami_2019}. In this section, it is shown that the same equation can also be derived 
for the case of steady axisymmetric vortex flow. Following a procedure analogous to that in~\cite{eckhardt_grossmann_lohse_2007}, 
the azimuthal momentum equation (\cref{eq:azim_steady_NV_visco_azi}) is averaged over the axial direction and one obtains
\begin{equation}\label{eq:azim_steady_avg}
      0 = -\overline{u\partial_r v} - \overline{w\partial_{z}v} -\frac{\overline{uv}}{r} + \frac{1}{Re}(\frac{1}{r}\partial_r(r\partial_r\overline{v}) - \frac{\overline{v}}{r^2})  + \frac{1-\beta}{Re}(\partial_r \overline{T_{r\theta}} + \frac{\overline{2T_{r\theta}}}{r}) , 
\end{equation}
\noindent Using integration by parts, $\overline{w\partial_{z}v} = -\overline{v\partial_{z}w}$, and the continuity equation,
$\partial_z w = -\partial_r u - u/r$, equation~\eqref{eq:azim_steady_avg} can be written as
\begin{equation}\label{eq:azim_steady_avg_replaced}
      0 = -\overline{u\partial_r v} - \overline{v\partial_r u} -\frac{2\overline{uv}}{r} + \frac{1}{Re}(\frac{1}{r}\partial_r(r\partial_r \overline{v}) - \frac{\overline{v}}{r^2})  + \frac{1-\beta}{Re}(\partial_r \overline{T_{r\theta}} + \frac{2\overline{T_{r\theta}}}{r}), 
\end{equation}
\noindent which can be rearranged into the form
\begin{equation}\label{eq:azim_steady_avg_rearr}
      0 = -r^{-2} \partial_r (r^2\overline{uv}) + r^{-2}\left[\frac{1}{Re}\partial_r (r^3 \partial_r \frac{\overline{v}}{r})\right]
      + r^{-2}\left[\frac{1-\beta}{Re}\partial_r (r^2\overline{T_{r\theta}})\right] 
\end{equation}
\noindent If we now multiply by $r^2$ and introduce the angular velocity $\omega = \frac{v}{r}$ the equation becomes
\begin{equation}\label{eq:azim_steady_avg_final}
      0 = \partial_r \left( r^{3} \left[ \overline{u\omega}  - \frac{1}{Re}\partial_r \overline{\omega}       -\frac{1-\beta}{Re}\partial_r \frac{\overline{T_{r\theta}}}{r}\right]\right)
\end{equation}

\noindent The equation above implies that the quantity 
\begin{equation}\label{eq:Jw}
      J^w = r^{3} \left[\overline{u\omega} - \frac{1}{Re}\partial_r \overline{\omega}
      -\frac{1-\beta}{Re}\partial_r \frac{\overline{T_{r\theta}}}{r}\right]
\end{equation}
\noindent does not change in the radial direction. Hence, it can be interpreted as a conserved current 
of the angular velocity across the annular gap. The three terms that appear in this equation correspond 
to the contributions of the different transport mechanisms (from left to right: transport due to convection, 
molecular diffusion and elastic stresses). Note that equation~\eqref{eq:Jw} is essentially the 
same as that derived for fully turbulent flow~\citep{song_teng_liu_ding_lu_khomami_2019}, with the only 
difference that while the average here is done only in space, time averaging is also needed in the 
turbulent case.\\

\section{Additional results about the variation of the angular momentum fluxes with $El$}\label{sec:appC}

\begin{figure}\setlength{\piclen}{\linewidth}
  \begin{center}
    \begin{tabular}{c}
      $(a)$ \\
      \includegraphics[width=\piclen]{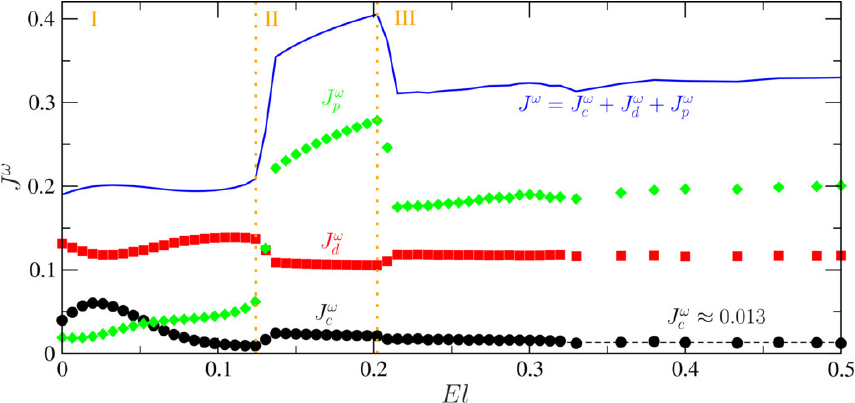} \\
      $(b)$ \\
      \includegraphics[width=\piclen]{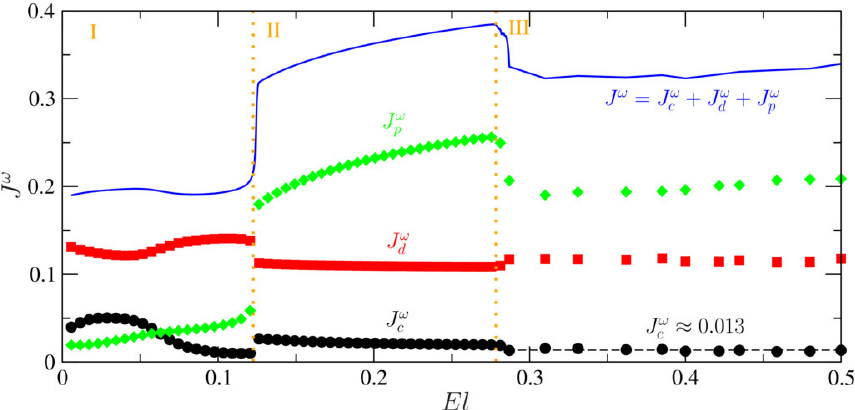} \\
      $(c)$ \\
      \includegraphics[width=\piclen]{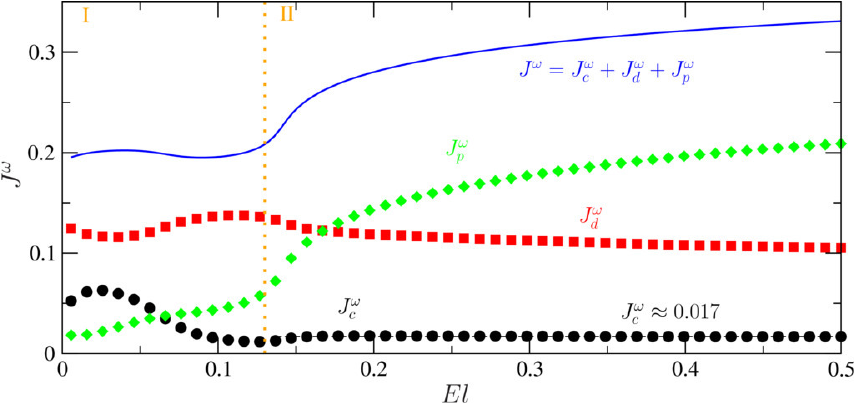} \\
    \end{tabular}
  \end{center}
  \caption{ (Color online)  Variation of the angular velocity current, $J^{\omega}$, and its components, $J^{\omega}_c$, $J^{\omega}_d$ and $J^{\omega}_p$, with $El$. Panels (a) and (b) correspond to simulations where the VMS regime takes place, whereas panels (c) exemplifies a case where this regime does not occur. More specifically, the data shown in the  
  panels (a) and (b) correspond to the simulations whose space time diagrams are illustrated in the panels (b) and (c) of the figure~\ref{fig:domain_size}, whereas the data shown in the panel (c) 
  corresponds to the simulation illustrated in the panel (a) of the figure~\ref{fig:L_effect_space_time}.}
  \label{fig:momentum_others} 
\end{figure}

\noindent The variation of the angular velocity current, $J^{\omega}$, with increasing $El$ obtained in all 
simulations where the VMS regime is found is analogous to that exemplified in the figure~\ref{fig:momentum} 
for the simulation presented in the section~\ref{sec:to_VMST}. Panels (a) and (b) of the figure~\ref{fig:momentum_others} 
show the variation of $J^{\omega}$ and its components ($J^{\omega}_c$, $J^{\omega}_d$ and $J^{\omega}_p$) with $El$ 
for the simulations illustrated in the panels (b) and (c) of the figure~\ref{fig:domain_size}.  It is clearly 
seen in these panels the existence of the three regimes described throughout the paper: the regime dominated 
by centrifugal effects (region I), the regime dominated by elastic effects where the flow 
pattern is steady (region II) and the regime characterized by spatio-temporal dynamics, including the VMS events 
(region III). The appearance of the latter regime is always accompanied by a slight decrease in the convective 
contribution, $J^{\omega}_c$ (the contribution of the vortices), which subsequently oscillates around a 
mean value indicated by a (black) dashed line in the figures. It is important to note that such mean value 
is nearly the same in all cases, $J^{\omega}_c \approx 0.013$, suggesting that at the Reynolds number at which the 
simulations are performed, $Re=95$, this may be the critical level for which the diwhirls fully decouple. 
For comparison, panels (c) in the figure~\ref{fig:momentum_others} shows the variation of $J^{\omega}$ with $El$ in 
a simulation where the VMS regime does not occur. More specifically, it corresponds to the simulation where 
$L = 70$, whose space time diagram is depicted in the panel (a) of the figure~\ref{fig:L_effect_space_time}.  
In this simulation, after the elastic instability sets in (zone II), the convective and diffusive contributions, 
$J^{\omega}_c$ and $J^{\omega}_d$, remain constant with increasing $El$, but the value at which of $J^{\omega}_c$ 
levels off ($J^{\omega}_c \approx 0.017$) is above that corresponding to the VMS regime. This observation is 
consistent with the hypothesis that diwhirls get fully decoupled only when their contribution to the angular 
momentum transport reaches a critical level. 

{\color{black}
\bibliography{local}

\begin{thebibliography}{61}
\expandafter\ifx\csname natexlab\endcsname\relax\def\natexlab#1{#1}\fi

\bibitem[Al-Mubaiyedh {\em et~al.\/}(1999)Al-Mubaiyedh, Sureshkumar \&
  Khomami]{AlSuKho99}
{\sc Al-Mubaiyedh, U.~A., Sureshkumar, R. \& Khomami, B.} 1999 Influence of
  energetics on the stability of viscoelastic {T}aylor–{C}ouette flow. {\em
  Physics of Fluids\/} {\bf 11}~(11), 3217--3226.

\bibitem[Andereck {\em et~al.\/}(1986)Andereck, Liu \&
  Swinney]{andereck_liu_swinney_1986}
{\sc Andereck, C.~D., Liu, S.~S. \& Swinney, H.~L.} 1986 Flow regimes in a
  circular {C}ouette system with independently rotating cylinders. {\em Journal
  of Fluid Mechanics\/} {\bf 164}, 155–183.

\bibitem[Aref(1983)]{Aref83}
{\sc Aref, H.} 1983 Integrable, chaotic, and turbulent vortex motion in
  two-dimensional flows. {\em Annual Review of Fluid Mechanics\/} {\bf 15}~(1),
  345--389.

\bibitem[Avila {\em et~al.\/}(2008)Avila, Grimes, Lopez \&
  Marques]{AvGriLoMa08}
{\sc Avila, M., Grimes, M., Lopez, J.~M. \& Marques, F.} 2008 Global endwall
  effects on centrifugally stable flows. {\em Physics of Fluids\/} {\bf
  20}~(10), 104104.

\bibitem[Batchelor(2000)]{batchelor_2000}
{\sc Batchelor, G.~K.} 2000 {\em An Introduction to Fluid Dynamics\/}.
  Cambridge University Press.

\bibitem[Baumert \& Muller(1995)]{baumert1995flow}
{\sc Baumert, B.~M. \& Muller, S.~J.} 1995 Flow visualization of the elastic
  {T}aylor-{C}ouette instability in {B}oger fluids. {\em Rheologica Acta\/}
  {\bf 34}~(2), 147--159.

\bibitem[Baumert \& Muller(1997)]{BauMu97}
{\sc Baumert, B.~M. \& Muller, S.~J.} 1997 Flow regimes in model viscoelastic
  fluids in a circular {C}ouette system with independently rotating cylinders.
  {\em Physics of Fluids\/} {\bf 9}~(3), 566--586.

\bibitem[Baumert \& Muller(1999)]{BAUMERT99}
{\sc Baumert, B.~M. \& Muller, S.~J.} 1999 Axisymmetric and non-axisymmetric
  elastic and inertio-elastic instabilities in {T}aylor–{C}ouette flow. {\em
  Journal of Non-Newtonian Fluid Mechanics\/} {\bf 83}~(1), 33--69.

\bibitem[Beris \& Dimitropoulos(1999)]{Beris99}
{\sc Beris, A.~N. \& Dimitropoulos, C.~D.} 1999 Pseudospectral simulation of
  turbulent viscoelastic channel flow. {\em Computer Methods in Applied
  Mechanics and Engineering\/} {\bf 180}~(3), 365 -- 392.

\bibitem[Bird {\em et~al.\/}(1980)Bird, Dotson \& Johnson]{Bird80}
{\sc Bird, R., Dotson, P. \& Johnson, N.} 1980 Polymer solution rheology based
  on a finitely extensible bead—spring chain model. {\em Journal of
  Non-Newtonian Fluid Mechanics\/} {\bf 7}~(2), 213 -- 235.

\bibitem[Cagney {\em et~al.\/}(2020)Cagney, Lacassagne \&
  Balabani]{cagney_lacassagne_balabani_2020}
{\sc Cagney, N., Lacassagne, T. \& Balabani, S.} 2020 {T}aylor–{C}ouette flow
  of polymer solutions with shear-thinning and viscoelastic rheology. {\em
  Journal of Fluid Mechanics\/} {\bf 905}, A28.

\bibitem[Coles(1965)]{coles65}
{\sc Coles, D.} 1965 Transition in circular {C}ouette flow. {\em Journal of
  Fluid Mechanics\/} {\bf 21}~(3), 385–425.

\bibitem[Crumeyrolle {\em et~al.\/}(2002)Crumeyrolle, Mutabazi \&
  Grisel]{CruMuGrise02}
{\sc Crumeyrolle, O., Mutabazi, I. \& Grisel, M.} 2002 Experimental study of
  inertioelastic {C}ouette–{T}aylor instability modes in dilute and
  semidilute polymer solutions. {\em Physics of Fluids\/} {\bf 14}~(5),
  1681--1688.

\bibitem[Czarny {\em et~al.\/}(2003)Czarny, Serre, Bontoux \&
  Lueptow]{CzSeBoLue03}
{\sc Czarny, O., Serre, E., Bontoux, P. \& Lueptow, R.~M.} 2003 Interaction
  between {E}kman pumping and the centrifugal instability in
  {T}aylor–{C}ouette flow. {\em Physics of Fluids\/} {\bf 15}~(2), 467--477.

\bibitem[Dallas {\em et~al.\/}(2010)Dallas, Vassilicos \&
  Hewitt]{dallas2010strong}
{\sc Dallas, V., Vassilicos, J.~C. \& Hewitt, G.~F.} 2010 Strong
  polymer-turbulence interactions in viscoelastic turbulent channel flow. {\em
  Physical Review E\/} {\bf 82}~(6), 066303.

\bibitem[Dessup {\em et~al.\/}(2018)Dessup, Tuckerman, Wesfreid, Barkley \&
  Willis]{DeTuWeBaWi18}
{\sc Dessup, T., Tuckerman, L.~S., Wesfreid, J.~E., Barkley, D. \& Willis,
  A.~P.} 2018 Self-sustaining process in {T}aylor-{C}ouette flow. {\em Phys.
  Rev. Fluids\/} {\bf 3}, 123902.

\bibitem[Dubief {\em et~al.\/}(2013)Dubief, Terrapon \& Soria]{Dubief13}
{\sc Dubief, Y., Terrapon, V.~E. \& Soria, J.} 2013 On the mechanism of
  elasto-inertial turbulence. {\em Physics of Fluids\/} {\bf 25}~(11), 110817.

\bibitem[Eckhardt {\em et~al.\/}(2007)Eckhardt, Grossmann \&
  Lohse]{eckhardt_grossmann_lohse_2007}
{\sc Eckhardt, B., Grossmann, S. \& Lohse, D.} 2007 Torque scaling in turbulent
  {T}aylor–{C}ouette flow between independently rotating cylinders. {\em
  Journal of Fluid Mechanics\/} {\bf 581}, 221–250.

\bibitem[Fenstermacher {\em et~al.\/}(1979)Fenstermacher, Swinney \&
  Gollub]{FensSwiGol79}
{\sc Fenstermacher, P.~R., Swinney, H.~L. \& Gollub, J.~P.} 1979 Dynamical
  instabilities and the transition to chaotic {T}aylor vortex flow. {\em
  Journal of Fluid Mechanics\/} {\bf 94}~(1), 103–128.

\bibitem[Fouxon \& Lebedev(2003)]{Foux03}
{\sc Fouxon, A. \& Lebedev, V.} 2003 Spectra of turbulence in dilute polymer
  solutions. {\em Physics of Fluids\/} {\bf 15}~(7), 2060--2072.

\bibitem[Gorman \& Swinney(1982)]{gorman_swinney_1982}
{\sc Gorman, M. \& Swinney, H.~L.} 1982 Spatial and temporal characteristics of
  modulated waves in the circular {C}ouette system. {\em Journal of Fluid
  Mechanics\/} {\bf 117}, 123–142.

\bibitem[Groisman \& Steinberg(1996)]{GroStei96}
{\sc Groisman, A. \& Steinberg, V.} 1996 {C}ouette-{T}aylor flow in a dilute
  polymer solution. {\em Phys. Rev. Lett.\/} {\bf 77}, 1480--1483.

\bibitem[Groisman \& Steinberg(1997)]{GroiStei97}
{\sc Groisman, A. \& Steinberg, V.} 1997 Solitary vortex pairs in viscoelastic
  {C}ouette flow. {\em Phys. Rev. Lett.\/} {\bf 78}, 1460--1463.

\bibitem[Groisman \& Steinberg(1998)]{GroiStei98}
{\sc Groisman, A. \& Steinberg, V.} 1998 Mechanism of elastic instability in
  {C}ouette flow of polymer solutions: Experiment. {\em Physics of Fluids\/}
  {\bf 10}~(10), 2451--2463.

\bibitem[Groisman \& Steinberg(2000)]{GroStei00}
{\sc Groisman, A. \& Steinberg, V.} 2000 Elastic turbulence in a polymer
  solution flow. {\em Nature\/} {\bf 405}, 53--55.

\bibitem[Groisman \& Steinberg(2004)]{GroStei04}
{\sc Groisman, A. \& Steinberg, V.} 2004 Elastic turbulence in curvilinear
  flows of polymer solutions. {\em New Journal of Physics\/} {\bf 6}.

\bibitem[Hegseth {\em et~al.\/}(1996)Hegseth, Baxter \& Andereck]{HegBaxAnd96}
{\sc Hegseth, J.~J., Baxter, G.~W. \& Andereck, C.~D.} 1996 Bifurcations from
  {T}aylor vortices between corotating concentric cylinders. {\em Phys. Rev.
  E\/} {\bf 53}, 507--521.

\bibitem[Jones(1981)]{jones_1981}
{\sc Jones, C.~A.} 1981 Nonlinear {T}aylor vortices and their stability. {\em
  Journal of Fluid Mechanics\/} {\bf 102}, 249–261.

\bibitem[Jones(1985)]{jones_1985}
{\sc Jones, C.~A.} 1985 The transition to wavy {T}aylor vortices. {\em Journal
  of Fluid Mechanics\/} {\bf 157}, 135–162.

\bibitem[King {\em et~al.\/}(1984)King, Lee, Swinney \&
  Marcus]{king_lee_swinney_marcus_1984}
{\sc King, G.~P., Lee, Y.~Li, W., Swinney, H.~L. \& Marcus, P.~S.} 1984 Wave
  speeds in wavy {T}aylor-vortex flow. {\em Journal of Fluid Mechanics\/} {\bf
  141}, 365–390.

\bibitem[Lacassagne {\em et~al.\/}(2021)Lacassagne, Cagney \&
  Balabani]{lacassagne_cagney_balabani_2021}
{\sc Lacassagne, T., Cagney, N. \& Balabani, S.} 2021 Shear-thinning mediation
  of elasto-inertial {T}aylor–{C}ouette flow. {\em Journal of Fluid
  Mechanics\/} {\bf 915}, A91.

\bibitem[Lacassagne {\em et~al.\/}(2020)Lacassagne, Cagney, Gillissen \&
  Balabani]{LaCaGiBa20}
{\sc Lacassagne, T., Cagney, N., Gillissen, J. J.~J. \& Balabani, S.} 2020
  Vortex merging and splitting: A route to elastoinertial turbulence in
  {T}aylor-{C}ouette flow. {\em Phys. Rev. Fluids\/} {\bf 5}, 113303.

\bibitem[Lange \& Eckhardt(2001)]{LaBru01}
{\sc Lange, M. \& Eckhardt, B.} 2001 Vortex pairs in viscoelastic
  {C}ouette-{T}aylor flow. {\em Phys. Rev. E\/} {\bf 64}, 027301.

\bibitem[Larson {\em et~al.\/}(1990)Larson, Shaqfeh \&
  Muller]{larson_shaqfeh_muller_1990}
{\sc Larson, R.~G., Shaqfeh, E. S.~G. \& Muller, S.~J.} 1990 A purely elastic
  instability in {T}aylor–{C}ouette flow. {\em Journal of Fluid Mechanics\/}
  {\bf 218}, 573–600.

\bibitem[Latrache {\em et~al.\/}(2016)Latrache, Abcha, Crumeyrolle \&
  Mutabazi]{LaAbCruMu16}
{\sc Latrache, N., Abcha, N., Crumeyrolle, O. \& Mutabazi, I.} 2016
  Defect-mediated turbulence in ribbons of viscoelastic {T}aylor-{C}ouette
  flow. {\em Phys. Rev. E\/} {\bf 93}, 043126.

\bibitem[Latrache {\em et~al.\/}(2012)Latrache, Crumeyrolle \&
  Mutabazi]{LaCruMu12}
{\sc Latrache, N., Crumeyrolle, O. \& Mutabazi, I.} 2012 Transition to
  turbulence in a flow of a shear-thinning viscoelastic solution in a
  {T}aylor-{C}ouette cell. {\em Phys. Rev. E\/} {\bf 86}, 056305.

\bibitem[Latrache \& Mutabazi(2021)]{latrache2021transition}
{\sc Latrache, N. \& Mutabazi, I.} 2021 Transition to turbulence via flame
  patterns in viscoelastic {T}aylor--{C}ouette flow. {\em The European Physical
  Journal E\/} {\bf 44}~(5), 1--15.

\bibitem[Liu \& Khomami(2013)]{liu_khomami_2013}
{\sc Liu, N. \& Khomami, B.} 2013 Elastically induced turbulence in
  {T}aylor–{C}ouette flow: direct numerical simulation and mechanistic
  insight. {\em Journal of Fluid Mechanics\/} {\bf 737}, R4.

\bibitem[Lopez \& Avila(2017)]{LoAv17}
{\sc Lopez, J.~M. \& Avila, M.} 2017 Boundary-layer turbulence in experiments
  on quasi-{K}eplerian flows. {\em Journal of Fluid Mechanics\/} {\bf 817},
  21–34.

\bibitem[Lopez {\em et~al.\/}(2019)Lopez, Choueiri \&
  Hof]{lopez_choueiri_hof_2019}
{\sc Lopez, J.~M., Choueiri, G.~H. \& Hof, B.} 2019 Dynamics of viscoelastic
  pipe flow at low {R}eynolds numbers in the maximum drag reduction limit. {\em
  Journal of Fluid Mechanics\/} {\bf 874}, 699–719.

\bibitem[L{\'o}pez {\em et~al.\/}(2020)L{\'o}pez, Feldmann, Rampp,
  Vela-Mart{\'\i}n, Shi \& Avila]{lopez2020nscouette}
{\sc L{\'o}pez, J.~M., Feldmann, D., Rampp, M., Vela-Mart{\'\i}n, A., Shi, L.
  \& Avila, M.} 2020 ns{C}ouette--a high-performance code for direct numerical
  simulations of turbulent {T}aylor--{C}ouette flow. {\em SoftwareX\/} {\bf
  11}, 100395.

\bibitem[Marcus(1984{\natexlab{{\em a\/}}})]{marcus_1984_1}
{\sc Marcus, P.~S.} 1984{\natexlab{{\em a\/}}} Simulation of {T}aylor-{C}ouette
  flow. part 1. numerical methods and comparison with experiment. {\em Journal
  of Fluid Mechanics\/} {\bf 146}, 45–64.

\bibitem[Marcus(1984{\natexlab{{\em b\/}}})]{marcus_1984_2}
{\sc Marcus, P.~S.} 1984{\natexlab{{\em b\/}}} Simulation of {T}aylor-{C}ouette
  flow. part 2. numerical results for wavy-vortex flow with one travelling
  wave. {\em Journal of Fluid Mechanics\/} {\bf 146}, 65–113.

\bibitem[Martinand {\em et~al.\/}(2014)Martinand, Serre \&
  Lueptow]{MaSerrLuep14}
{\sc Martinand, D., Serre, E. \& Lueptow, R.~M.} 2014 Mechanisms for the
  transition to waviness for {T}aylor vortices. {\em Physics of Fluids\/} {\bf
  26}~(9), 094102.

\bibitem[Muller {\em et~al.\/}(1989)Muller, Larson \&
  Shaqfeh]{muller1989purely}
{\sc Muller, S.~J., Larson, R.~G. \& Shaqfeh, E.~S.} 1989 A purely elastic
  transition in {T}aylor-{C}ouette flow. {\em Rheologica Acta\/} {\bf 28}~(6),
  499--503.

\bibitem[Ruelle \& Takens(1971)]{RueTa71}
{\sc Ruelle, D. \& Takens, F.} 1971 On the nature of turbulence. {\em Commun.
  Math. Phys.\/} {\bf 20}, 167--192.

\bibitem[Samanta {\em et~al.\/}(2013)Samanta, Dubief, Holzner, Sch{\"a}fer,
  Morozov, Wagner \& Hof]{Samanta13}
{\sc Samanta, D., Dubief, Y., Holzner, M., Sch{\"a}fer, C., Morozov, A.~N.,
  Wagner, C. \& Hof, B.} 2013 Elasto-inertial turbulence. {\em Proceedings of
  the National Academy of Sciences\/} {\bf 110}~(26), 10557--10562.

\bibitem[Shaqfeh(1996)]{Sha96}
{\sc Shaqfeh, E. S.~G.} 1996 Purely elastic instabilities in viscometric flows.
  {\em Annual Review of Fluid Mechanics\/} {\bf 28}~(1), 129--185.

\bibitem[Shi {\em et~al.\/}(2015)Shi, Rampp, Hof \& Avila]{shi2015hybrid}
{\sc Shi, L., Rampp, M., Hof, B. \& Avila, M.} 2015 A hybrid mpi-openmp
  parallel implementation for pseudospectral simulations with application to
  {T}aylor--{C}ouette flow. {\em Computers \& Fluids\/} {\bf 106}, 1--11.

\bibitem[Song {\em et~al.\/}(2021{\natexlab{{\em a\/}}})Song, Lin, Liu, Lu \&
  Khomami]{song_lin_liu_lu_khomami_2021}
{\sc Song, J., Lin, F., Liu, N., Lu, X.-Y. \& Khomami, B.} 2021{\natexlab{{\em
  a\/}}} Direct numerical simulation of inertio-elastic turbulent
  {T}aylor–{C}ouette flow. {\em Journal of Fluid Mechanics\/} {\bf 926}, A37.

\bibitem[Song {\em et~al.\/}(2019)Song, Teng, Liu, Ding, Lu \&
  Khomami]{song_teng_liu_ding_lu_khomami_2019}
{\sc Song, J., Teng, H., Liu, N., Ding, H., Lu, X.-Y. \& Khomami, B.} 2019 The
  correspondence between drag enhancement and vortical structures in turbulent
  {T}aylor–{C}ouette flows with polymer additives: a study of curvature
  dependence. {\em Journal of Fluid Mechanics\/} {\bf 881}, 602–616.

\bibitem[Song {\em et~al.\/}(2021{\natexlab{{\em b\/}}})Song, Wan, Liu, Lu \&
  Khomami]{song_wan_liu_lu_khomami_2021}
{\sc Song, J., Wan, Z.-H., Liu, N., Lu, X.-Y. \& Khomami, B.}
  2021{\natexlab{{\em b\/}}} A reverse transition route from inertial to
  elasticity-dominated turbulence in viscoelastic {T}aylor–{C}ouette flow.
  {\em Journal of Fluid Mechanics\/} {\bf 927}, A10.

\bibitem[Taylor(1923)]{Taylor23}
{\sc Taylor, G. I.~S.} 1923 Stability of a viscous liquid contained between two
  rotating cylinders. {\em Philosophical Transactions of the Royal Society A\/}
  {\bf 223}, 289--343.

\bibitem[Thomas {\em et~al.\/}(2009)Thomas, Khomami \&
  Sureshkumar]{thomas_khomami_sureshkumar_2009}
{\sc Thomas, D.~G., Khomami, B. \& Sureshkumar, R.} 2009 Nonlinear dynamics of
  viscoelastic {T}aylor–{C}ouette flow: effect of elasticity on pattern
  selection, molecular conformation and drag. {\em Journal of Fluid
  Mechanics\/} {\bf 620}, 353–382.

\bibitem[Thomas {\em et~al.\/}(2006)Thomas, Sureshkumar \&
  Khomami]{ThoSurKho06}
{\sc Thomas, D.~G., Sureshkumar, R. \& Khomami, B.} 2006 Pattern formation in
  {T}aylor-{C}ouette flow of dilute polymer solutions: Dynamical simulations
  and mechanism. {\em Phys. Rev. Lett.\/} {\bf 97}, 054501.

\bibitem[Wereley \& Lueptow(1998)]{wereley_lueptow_1998}
{\sc Wereley, S.~T. \& Lueptow, R.~M.} 1998 Spatio-temporal character of
  non-wavy and wavy {T}aylor–{C}ouette flow. {\em Journal of Fluid
  Mechanics\/} {\bf 364}, 59–80.

\bibitem[Willis(2017)]{openpipeflow}
{\sc Willis, A.~P.} 2017 The openpipeflow {N}avier–{S}tokes solver. {\em
  SoftwareX\/} {\bf 6}, 124 -- 127.

\bibitem[Xi \& Graham(2010)]{XiGra10b}
{\sc Xi, L. \& Graham, M.~D.} 2010 Turbulent drag reduction and multistage
  transitions in viscoelastic minimal flow units. {\em J.\,Fluid Mech.\/} {\bf
  647}, 421–452.

\bibitem[Yamani {\em et~al.\/}(2021)Yamani, Keshavarz, Raj, Zaki, McKinley \&
  Bischofberger]{Ya21}
{\sc Yamani, S., Keshavarz, B., Raj, Y., Zaki, T.~A., McKinley, G.~H. \&
  Bischofberger, I.} 2021 Spectral universality of elastoinertial turbulence.
  {\em Phys. Rev. Lett.\/} {\bf 127}, 074501.

\bibitem[Zhu {\em et~al.\/}(2022)Zhu, Song, Lin, Liu, Lu \&
  Khomami]{zhu_song_lin_liu_lu_khomami_2022}
{\sc Zhu, Y., Song, J., Lin, F., Liu, N., Lu, X. \& Khomami, B.} 2022
  Relaminarization of spanwise-rotating viscoelastic plane {C}ouette flow via a
  transition sequence from a drag-reduced inertial to a drag-enhanced
  elasto-inertial turbulent flow. {\em Journal of Fluid Mechanics\/} {\bf 931},
  R7.

\bibitem[Zhu {\em et~al.\/}(2020)Zhu, Song, Liu, Lu \&
  Khomami]{zhu_song_liu_lu_khomami_2020}
{\sc Zhu, Y., Song, J., Liu, N., Lu, X. \& Khomami, B.} 2020 Polymer-induced
  flow relaminarization and drag enhancement in spanwise-rotating plane
  {C}ouette flow. {\em Journal of Fluid Mechanics\/} {\bf 905}, A19.

\end{thebibliography}
\bibliographystyle{jfm}
}
\end{document}